\documentclass[aps,eqsecnum,preprint,floats,epsf,epsfig,nofootinbib]{revtex4}
\usepackage{graphicx}
\usepackage{epsfig}

\begin{document}
\def\be{\begin{eqnarray}}
\def\en{\end{eqnarray}}
\def\non{\nonumber}
\def\la{\langle}
\def\ra{\rangle}
\def\vp{\varepsilon}
\def\hep{\hat{\varepsilon}}
\def\a{{\cal A}}
\def\B{{\cal B}}
\def\c{{\cal C}}
\def\d{{\cal D}}
\def\e{{\cal E}}
\def\P{{\cal P}}
\def\t{{\cal T}}
\def\up{\uparrow}
\def\dw{\downarrow}
\def\3bar{{\bf \bar 3}}
\def\6bar{{\bf \bar 6}}
\def\10bar{{\bf \ov{10}}}
\def\ov{\overline}
\def\Lqcd{{\Lambda_{\rm QCD}}}
\def\pr{{Phys. Rev.}~}
\def\prl{{ Phys. Rev. Lett.}~}
\def\pl{{ Phys. Lett.}~}
\def\np{{ Nucl. Phys.}~}
\def\zp{{ Z. Phys.}~}
\def\lsim{ {\ \lower-1.2pt\vbox{\hbox{\rlap{$<$}\lower5pt\vbox{\hbox{$\sim$}
}}}\ } }
\def\gsim{ {\ \lower-1.2pt\vbox{\hbox{\rlap{$>$}\lower5pt\vbox{\hbox{$\sim$}
}}}\ } }

\font\el=cmbx10 scaled \magstep2{\obeylines\hfill June, 2004}

\vskip 1.5 cm

\centerline{\large\bf Light-Front Approach for Pentaquark Strong
Decays}
\bigskip
\centerline{\bf Hai-Yang Cheng and  Chun-Khiang Chua}
\medskip
\centerline{Institute of Physics, Academia Sinica}
\centerline{Taipei, Taiwan 115, Republic of China}
\medskip

\bigskip
\bigskip
\centerline{\bf Abstract}
 \small
Assuming the two diquark structure for the pentaquark state as
advocated in the Jaffe-Wilczek model, we study the strong decays
of light and heavy parity-even pentaquark states using the
light-front quark model in conjunction with the spectator
approximation. The narrowness of the $\Theta^+$ width is ascribed
to the $p$-wave configuration of the diquark pair. Taking the
$\Theta^+$ width as a benchmark, we estimate the rates of the
strong decays $\Xi^{--}_{3/2}\to\Xi^-\pi^-,\Sigma^-K^-$,
$\Sigma_{5c}^0\to D_s^- p,\,D_{s0}^{*-}p$ and $\Xi_{5c}^0\to
D_s^-\Sigma^+,D_{s0}^{*-}\Sigma^+$ with $\Sigma_{5c},\Xi_{5c}$
being antisextet charmed pentaquarks and $D_{s0}^*$ a scalar
strange charmed meson. The ratio of $\Gamma(\P_c\to{\cal B}
D^*_{s0})/\Gamma(\P_c\to{\cal B} D_s)$ is very useful for
verifying the parity of the antisextet charmed pentaquark $\P_c$.
It is expected to be of order unity for an even parity $\P_c$ and
much less than one for an odd parity pentaquark.

\bigskip

\eject
\section{Introduction}

The recent discovery of an exotic $\Theta^+$ baryon with $S=+1$ by
LEPS at SPring-8 \cite{LEPS}, subsequently confirmed by many other
groups
\cite{Diana,CLAS1,Saphir,ITEP,CLAS2,Hermes,SVD,COSY,ZEUS,Aslanyan},
led to a renewed interest in hadron spectroscopy and promoted a
re-examination of the QCD implications for exotic hadrons. The
mass of the $\Theta^+$ is of order 1535 MeV and its width is less
than 10 MeV from direct observations and can be as narrow as 1 MeV
from the analysis of $K$-deuteron scattering data \cite{width} and
it is most likely an isosinglet. The $I=3/2$ exotic pentaquark
$\Xi^{--}_{3/2}$ with a mass of $1862\pm 2$ MeV and a width
smaller than 18 MeV was observed by NA49 \cite{NA49} (see
also~\cite{Fischer} for a critical discussion). In spite of the
confirmation of the $\Theta^+$ from several experiments, all
current experimental signals are weak and the significance is only
of 4--6 standard deviations. Indeed, there exist several null
results for the pentaquark search from
\cite{BES,HeraB,Phenix,ALEPH,DELPHI}. An effort for understanding
why the $\Theta^+$ is seen in some experiments but not in others
has been made in \cite{KLNst}. The pentaquark candidate signals
must be established beyond any doubt by increasing the
experimental statistics.

The $\Theta^+$ mass is expected to be of order 1900 MeV for an
$s$-wave ground state with odd parity and 2200 MeV for a $p$-wave
state with even parity in the conventional uncorrelated quark
model. The width is at least of order several hundred MeV as the
strong decay $\Theta^+\to KN$ is Okubo-Zweig-Iizuka (OZI)
super-allowed. Therefore, within the naive uncorrelated quark
model one cannot understand why $\Theta^+$ is anomalously light
and why its width is so narrow. This hints a possible correlation
among various quarks; two or three quarks could form a cluster.
Several quark cluster models have been proposed in the past
\cite{JW,KL,SZ}. For example, Jaffe and Wilczek \cite{JW}
advocated a two diquark picture in which the $\Theta^+$ is a bound
state of an $\bar s$ quark with two $(ud)$ diquarks. The diquark
is a highly correlated spin-zero object and is in a flavor
anti-triplet and color anti-triplet state. The parity of
$\Theta^+$ is flipped from the negative, as expected in the naive
quark model, to the positive owing to the diquark correlation. The
even parity of the $\Theta^+$ is in agreement with the prediction
of the chiral soliton model \cite{DPP}.  Note that two of previous
lattice calculations imply a negative parity for the
$\Theta^+$~\cite{Csikor,Sasaki}. However, based on the
Jaffe-Wilczek picture to construct the interpolating operators, a
recent quenched lattice QCD calculation with exact chiral symmetry
yields a positive parity for the pentaquark states $\Theta^+$,
$\Xi^{--}_{3/2}$ \cite{Chiu} and for charmed pentaquarks to be
discussed below \cite{Chiu2}.

It is natural to consider the heavy flavor analogs $\Theta_c^0$
and $\Theta_b^+$ of $\Theta^+$ by replacing the $\bar s$ quark in
$\Theta^+$ by the heavy antiquark $\bar c$ and $\bar b$,
respectively. Whether the mass of the heavy pentaquark state is
above or below the strong-decay threshold has been quite
controversial. Very recently, a narrow resonance in $D^{*-}p$ and
$D^{*+}\bar p$ invariant mass distributions was reported by the H1
Collaboration~\cite{H1}. It has a mass of $3099\pm3\pm5$ MeV and a
Gaussian width of $12\pm3$ MeV and can be identified with the spin
1/2 or 3/2 charmed pentaquark baryon. However, there are also
several null results reported by ZEUS \cite{noThetac}, ALEPH
\cite{ALEPH} and FOCUS \cite{FOCUS}. Although the state observed
by H1 is about 300 MeV higher than the $DN$ threshold, it is
possible that the observed H1 pentaquark is a chiral partner of
the yet undiscovered ground state $\Theta_c^0$ with opposite
parity and a mass of order 2700 MeV as implied by several model
estimates \cite{Nowak}. The latter pentaquark can be discovered
only through its weak decay~\cite{CCH04}. Note that the
theoretical estimates of $\Theta_c$ mass are controversial even
within the Jaffe-Wilczek picture: The original estimate made by
Jaffe and Wilczek is below the $D^{(*)}p$ threshold~\cite{JW},
while other calculations in \cite{Cheung,Chiu2} that take into
account hyperfine interactions between the anti-charmed quark and
the two diquarks yield a charmed pentaquark mass above the
strong-decay threshold. The latter is also preferred by a recent
QCD sum rule calculation \cite{KLO}. Given the situation, it is
therefore interesting to consider the strong decays of charmed
pentaquarks as well.

In the Jaffe-Wilczek model, there exist parity-even antisextet and
parity-odd triplet heavy pentaquarks containing a single heavy
antiquark $\bar c$ or $\bar b$ and they are all truly exotic. The
heavy pentaquark baryons in the ${\bf 3}_f$ representation are
lighter than the $\6bar_f$ ones due to the lack of orbital
excitation and therefore may be stable against strong decays
\cite{Stewart,Chiu2}. Consequently, it becomes important to study
the weak decays of triplet heavy
pentaquarks~\cite{CCH04,Leibovich}.
In \cite{CCH04} we have employed the relativistic light-front (LF)
approach to study the heavy pentaquark weak decays. It is found
that the weak transition form factors thus obtained are consistent
with heavy quark symmetry~\cite{{IW89,IW91,Georgi,Yan92}}.

The light-front model allows us to study the transition form
factors and their momentum dependence. Furthermore, large
relativistic effects which may manifest near the maximum large
recoil, i.e. $q^2=0$, are properly taken into account in the
light-front framework. In this work we shall extend the formalism
to pentaquark strong decays. The strong decays of pentaquarks have
been studied in
\cite{Carlson:2003pn,Jaffe:2003ci,He,Mehen,Liu,Melikhov} and they
can be classified into
 \be
 (a) && \P(\overline{\bf 10})\to \B({\bf 8})+M,  \non \\
 (b) && \P_Q(\6bar)\to \B({\bf 8})+M_Q,
 \en
where $\P_{(Q)}$ denotes a generic (heavy) pentaquark baryon,
$\B({\bf 8})$ stands for the usual octet baryon made of three
quarks and $M_{(Q)}$ is a (heavy) meson. Examples  are
$\Theta^+\to pK^0$, $\Xi^{--}\to\Xi^-\pi^-,\Sigma^-K^-$,
$\Sigma^0_{5c}\to D_s^- p$, $\Xi_{5c}^0\to \ov
D^0\Xi^0,D_s^-\Sigma^+,D_{s0}^{*-}\Sigma^+$. Of course, whether
the above-mentioned strong decays are kinematically allowed or not
depends on the (heavy) pentaquark masses. It is interesting to
understand why the $\Theta^+$ width is much smaller than a typical
strong decay width.
We find that the narrowness of the $\Theta^+$ width is most likely
ascribed to the $p$-wave configuration of its constituent diquark
pair.
It is important to note that the scalar charmed meson $D_{s0}^*$
is experimentally found to have a mass of order 2317 MeV
\cite{BaBarDs,BelleDs}, which is considerably lighter than
expected from potential models \cite{QM}. Therefore, the
$D_{s0}^*{\cal B}$ threshold is not far from the $D_s{\cal B}$
one,  rendering the study of the decay $\P_c\to D_{s0}^*{\cal B}$
interesting. Indeed, since the parities of $D_{s0}^*$ and $\P_c$
(in the Jeffe-Wilczek model) are the same, the $D_{s0}^*{\cal B}$
final state can be in a $s$-wave configuration. Thus it is not
subject to a suppression near the threshold and hence can have a
sizable decay rate compared to $\P_c\to D_s{\cal B}$. This could
be useful for measuring the parity of $\P_c$.

The layout of the present paper is organized as follows. In Sec.
II we present a study of the pentaquark transitions within the
light-front quark model and derive the analytic expressions for
form factors.
Numerical results for form factors and examples of strong decays
of light and heavy pentaquark baryons are worked out in Sec. IV.
Conclusion is given in Sec. V followed by an Appendix devoted to
various baryon and meson Clebsch-Gordan coefficients.

\section{Formalism of a light-front model for pentaquarks}

In this section we shall focus on the hadronic strong decays of
light and heavy pentaquarks  within the light-front approach and
the Jaffe-Wilczek model.  In this study we need to use pentaquark
and meson vertex functions. We shall consider the mesonic case
first as it is simpler. Readers who are not interested in the
technical details of vertex functions can skip directly to Sec. II
B.

\subsection{Vertex functions in the light-front approach}

\subsubsection{Vertex functions for mesons}

In the conventional light-front approach, a meson bound state
consisting of a quark $q_1$ and an antiquark $\bar q_2$ with the
total momentum $P$ and spin $J$ can be written as (see, for
example, \cite{Cheng97} for odd-parity and \cite{CCH} for
even-parity mesons)
\begin{eqnarray}
        |M(P,\, ^{2S+1}L_J, J_z)\rangle
                &=& \int \{d^3p_1\}\{d^3p_2\} ~2(2\pi)^3 \delta^3(
                \tilde P -\tilde p_1-\tilde p_2)~\nonumber\\
        &\times& \sum_{\lambda_1,\lambda_2,\alpha,a,b}
                \Psi^{JJ_z}_{LS}(x_1,x_2,k_{1\bot},k_{2\bot})~M^b_a
                |(q^c)^{a\alpha}(p_1,\lambda_1) (\bar q^c)_{b\alpha}(p_2,\lambda_2)\rangle,
 \label{eq:lfmbs}
\end{eqnarray}
where $a,b$ are flavor indices~\footnote{Note that we use the
field convention instead of the particle convention to denote the
quantum numbers of the state, i.e. the state quantum number is
defined according to the field creating the state.}, $\alpha$ is
the color index, $M^b_a$ is a normalized matrix element
characterizing the meson $SU_f(3)$ quantum number (see the
Appendix for details), $p_1,p_2$ are the on-mass-shell light-front
momenta,
\begin{equation}
        \tilde p=(p^+, p_\bot)~, \quad p_\bot = (p^1, p^2)~,
                \quad p^- = {m^2+p_\bot^2\over p^+},
\end{equation}
and
\begin{eqnarray}
        &&\{d^3p\} \equiv {dp^+d^2p_\bot\over 2(2\pi)^3}, \nonumber \\
        &&|q^c(p_1,\lambda_1)\bar q^c(p_2,\lambda_2)\rangle
        = b^\dagger_{\lambda_1}(p_1)d^\dagger_{\lambda_2}(p_2)|0\rangle,\\
        &&\{b_{\lambda'}(p'),b_{\lambda}^\dagger(p)\} =
        \{d_{\lambda'}(p'),d_{\lambda}^\dagger(p)\} =
        2(2\pi)^3~\delta^3(\tilde p'-\tilde p)~\delta_{\lambda'\lambda}.
                \nonumber
\end{eqnarray}
Note that we use the charge conjugated fields for quarks. For
example, we shall use $|c^c(p_1,\lambda_1) \bar
d^c(p_2,\lambda_2)\ra$ in Eq.~(\ref{eq:lfmbs}) for the $D^-$
meson. The reason for using the charged conjugated field will
become clear later.

In terms of the light-front relative momentum variables $(x,
p_\bot)$ defined by
\begin{eqnarray}
        && p^+_1=x_1 P^{+}, \quad p^+_2=x_2 P^{+}, \quad x_1+x_2=1, \nonumber \\
        && p_{1\bot}=x_1 P_\bot+k_{1\bot}, \quad p_{2\bot}=x_2
                P_\bot+k_{2\bot},\quad k_\bot=k_{1\bot}=-k_{2\bot},
\end{eqnarray}
the momentum-space wave-function $\Psi^{JJ_z}_{LS}$ for a
$^{2S+1}L_J$ meson can be expressed as
\begin{equation}
        \Psi^{ JJ_z}_{LS}(x_1,x_2,k_{1\bot},k_{2\bot})
                = \frac{1}{\sqrt N_c}\la L S; L_z S_z|L S;J J_z\ra
                  R^{SS_z}_{\lambda_1\lambda_2}(x,k_\bot)~ \psi_{LL_z}(x, k_\bot),
\end{equation}
where $x\equiv x_2$, $\varphi_{LL_z}(x,k_\bot)$ describes the
momentum distribution of the constituent quarks in the bound state
with the orbital angular momentum $L$, $\la L S; L_z S_z|L S;J
J_z\ra$ is the corresponding Clebsch-Gordan coefficient and
$R^{SS_z}_{\lambda_1\lambda_2}$ constructs a state of definite
spin ($S,S_z$) out of light-front helicity ($\lambda_1,\lambda_2$)
eigenstates.  Explicitly~\cite{Jaus90,deAraujo:1999cr},
\begin{equation}
        R^{SS_z}_{\lambda_1 \lambda_2}(x,k_\bot)
              =\sum_{s_1,s_2} \langle \lambda_1|
                {\cal R}_M^\dagger(1-x,k_\bot, m_1)|s_1\rangle
                \langle \lambda_2|{\cal R}_M^\dagger(x,-k_\bot, m_2)
                |s_2\rangle
                \left\langle \frac{1}{2}\,\frac{1}{2};s_1
                s_2|\frac{1}{2}\frac{1}{2};SS_z\right\rangle,
\end{equation}
where $|s_i\rangle$ are the usual Pauli spinors, and ${\cal R}_M$
is the Melosh transformation
operator~\cite{Jaus90,deAraujo:1999cr}:
 \be
        \la s|{\cal R}_M (x,k_\bot,m_i)|\lambda\ra
        &=&\frac{\bar
        u_D(k_i,s) u(k_i,\lambda)}{2 m_i}=-\frac{\bar
        v(k_i,\lambda) v_D(k_i,s)}{2 m_i}
        \non\\
        &=&\frac{(m_i+x_i \bar M_0)\delta_{s\lambda}
                      +i\vec \sigma_{s\lambda}\cdot\vec k_\bot \times
                      \vec                n}
                {\sqrt{(m_i+x_i M_0)^2 + k^{2}_\bot}},
 \label{eq:Melosh}
 \en
with $u_{(D)}$, a Dirac spinor in the light-front (instant) form
which has the expression
 \be
 u_D(k,s)=\frac{\not\!k+m}{\sqrt{k^0+m}}
 \left(\begin{array}{c}
       \chi_s\\
       0
       \end{array}
 \right),
 \quad
  u(k,\lambda)=\frac{\not\!k+m}{\sqrt{2 k^+}}\gamma^+\gamma^0
 \left(\begin{array}{c}
       \chi_\lambda\\
       0
       \end{array}
 \right),
 \label{eq:u}
 \en
in the Dirac representation, $\vec n = (0,0,1)$, a unit vector in
the $z$-direction, and
 \be
 M_0^2={m_1^2+k^2_\bot\over x_1}+{m_2^2+k^2_\bot\over x_2}.
 \label{eq:M02q}
 \en
Note that
 $u_D(p,s)=u(p,\lambda) \la \lambda|{\cal R}^\dagger_M|s\ra$
and, consequently, the state $|q(p,\lambda)\ra \la \lambda|{\cal
R}^\dagger_M|s\ra$ transforms like $|q(p,s)\ra$ under rotation,
i.e. its transformation does not depend on its momentum. A crucial
feature of the light-front formulation of a bound state, such as
the one shown in Eq.~(\ref{eq:lfmbs}), is the frame-independence
of the light-front wave function~\cite{Brodsky:1997de,Jaus90}.
Namely, the hadron can be boosted to any (physical) ($P^+$,
$P_\bot$) without affecting the internal variables ($x$, $k_{\bot
}$) of the wave function, which is certainly not the case in the
instant-form formulation.

In practice it is more convenient to use the covariant form for
$R^{SS_z}_{\lambda_1\lambda_2}$ \cite{Jaus91}:
\begin{equation}
        R^{SS_z}_{\lambda_1\lambda_2}(x,k_\bot)
                =\frac{1}{\sqrt2~{\widetilde M_0}(M_0+m_1+m_2)}
        ~\bar u(p_1,\lambda_1)(\not\!\bar P+M_0)\Gamma\,v(p_2,\lambda_2),
        \label{covariant}
\end{equation}
with
\begin{eqnarray}
        &&\widetilde M_0\equiv\sqrt{M_0^2-(m_1-m_2)^2}, \non \\
        &&\bar P\equiv p_1+p_2, \non \\
        &&\varepsilon^\mu(\bar P,\pm 1) =
                \left[{2\over P^+} \vec \varepsilon_\bot (\pm 1) \cdot
                \vec P_\bot,\,0,\,\vec \varepsilon_\bot (\pm 1)\right],
                \quad \vec \varepsilon_\bot
                (\pm 1)=\mp(1,\pm i)/\sqrt{2}, \nonumber\\
        &&\varepsilon^\mu(\bar P,0)={1\over M_0}\left({-M_0^2+P_\bot^2\over
                P^+},P^+,P_\bot\right).   \label{polcom}
\end{eqnarray}
For the pseudoscalar and vector mesons, we have
\begin{eqnarray} \label{eq:Pvertex}
        &&\Gamma_P=\gamma_5 \qquad\quad ({\rm pseudoscalar},L=0, S=0), \non \\
        &&\Gamma_V=-\not{\! \varepsilon}(\bar P, S_z) \qquad\quad~ ({\rm vector},L=0,
        S=1),
\end{eqnarray}
where
\begin{eqnarray}
 M_0=e_1+e_2,\quad
 e_i =\sqrt{m^{2}_i+k^{2}_\bot+k^{2}_z},\quad
 k_z=\frac{x_1 M_0}{2}-\frac{m_1^2+k^{2}_\bot}{2 x_1 M_0}.
 \label{eq:internal2q}
 \end{eqnarray}
Applying equations of motion on spinors to Eq.~(\ref{covariant})
leads to
 \be
 \bar u(p_1)(\not\!\bar
 P+M_0)\gamma_5\,v(p_2)&=&(M_0+m_1+m_2) \bar u(p_1)\gamma_5\,v(p_2),
 \non\\
\bar u(p_1)(\not\!\bar
 P+M_0)\not{\! \varepsilon}\,v(p_2)&=&\bar u(p_1)[(M_0+m_1+m_2) \not{\!
 \varepsilon}-\varepsilon\cdot (p_1-p_2)]\,v(p_2),
 \label{eq:eom}
 \en
and $R^{SS_z}_{\lambda_1\lambda_2}$ is reduced to a more familiar
form~\cite{Jaus91}.
It is, however, more convenient to use the form shown in
Eq.~(\ref{covariant}) when extending to the $p$-wave meson case.
Two remarks are in order. First, $p_1+p_2$ is not equal to the
meson's four-momentum in the conventional LF approach as both the
quark and antiquark are on-shell. Second, the longitudinal
polarization 4-vector $\varepsilon^\mu(\bar P,0)$ given above is
not exactly the same as that of the vector meson and we have
$\varepsilon(\bar P,S_z)\cdot\bar P=0$. We normalize the meson
state as
\begin{equation}
        \langle M(P',J',J'_z)|M(P,J,J_z)\rangle = 2(2\pi)^3 P^+
        \delta^3(\tilde P'- \tilde P)\delta_{J'J}\delta_{J'_z J_z}~,
\label{wavenor}
\end{equation}
so that
\begin{equation}
        \int {dx\,d^2k_\bot\over 2(2\pi)^3}~\psi^{*}_{L^\prime L^\prime_z}(x,k_\bot)
                                                   \psi_{LL_z}(x,k_\bot)
        =\delta_{L^\prime L}~\delta_{L^\prime_z L_z}.
\label{momnor}
\end{equation}
Explicitly, we have
 \be
   \psi_{LL_z}(x,k_\bot)&=&\sqrt\frac{dk_{z}}{dx}~\varphi_{LL_z}(x,k_\bot),\qquad
   \frac{dk_{z}}{dx}=\frac{e_1 e_2}{x_1 x_2 M_0},
 \non\\
  \varphi_{00}(x, k_\bot)&=&\varphi(\vec k,\beta),\quad
  \varphi_{1m}(x, k_\bot)=k_{m} \varphi_p(\vec k,\beta),
  \label{eq:phi}
 \en
where $k_m=-\vec\varepsilon(m)\cdot \vec k=\varepsilon(\bar
P,m)\cdot (p_1-p_2)/2$, or explicitly $k_{m=\pm1}=\pm(k_{\bot
x}\pm i k_{\bot y})/\sqrt2$, $k_{m=0}=-k_{z}$ are proportional to
the spherical harmonics $Y_{1m}$ in the momentum space, and
$\varphi$, $\varphi_p$ are the distribution amplitudes of $s$-wave
and $p$-wave mesons, respectively. There are several popular
phenomenological light-front wave functions that have been
employed to describe various hadronic structures in the
literature. For a Gaussian-like wave function, one has
\cite{Cheng97,CCH}
\begin{eqnarray} \label{eq:Gauss}
 \varphi(\vec k,\beta)
    &=&4 \left({\pi\over{\beta^{2}}}\right)^{3\over{4}}
               ~{\rm exp}
               \left(-{k^2_z+k^2_\bot\over{2
               \beta^2}}\right),\quad
    \varphi_p(\vec k,\beta)=\sqrt{2\over{\beta^2}}~\varphi(\vec k,\beta).
 \label{eq:wavefn}
\end{eqnarray}
The parameter $\beta$ is expected to be of order $\Lambda_{\rm
QCD}$ and will be specified later.

It is straightforward to obtain~\cite{Jaus91,CCH04}
 \be
        iR^{00}_{\lambda_1\lambda_2}(x,k_\bot)
                &=&\frac{i}{\sqrt2~{\widetilde M_0}}
        ~\bar u(p_1,\lambda_1)\gamma_5\,v(p_2,\lambda_2),
 \non\\
 \la 1 S; L_z S_z|1 S;0 0\ra\, \varepsilon(\bar P, L_z)\cdot\frac{p_1-p_2}{2} \,R^{SS_z}_{\lambda_1\lambda_2}(x,k_\bot)
                &=&-\frac{\widetilde
                     M_0}{2\sqrt6~M_0}
        \bar u(p_1,\lambda_1) v(p_2,\lambda_2),
         \label{covariantp}
 \en
where $\la 1 S; L_z S_z|1 S;0 0\ra\, \varepsilon_\mu(\bar P, L_z)
\varepsilon_\nu(\bar P, S_z)=-\varepsilon_\mu^*(\bar P,
S_z)\varepsilon_\nu(\bar P, S_z)/\sqrt{3}$ have been made. Note
that an overall phase $i$ is assigned to the $^1S_0$ state to
match the usual phase convention. Putting everything together, we
have
\begin{eqnarray}
        |M(P,\, S^{\!\!\!\!\!\!\!\!1}_0, 0)\rangle
                &=&i\int \{d^3p_1\}\{d^3p_2\} ~2(2\pi)^3 \delta^3(
                \tilde P -\tilde p_1-\tilde p_2)~\nonumber\\
        &&\times \sum_{\lambda_1,\lambda_2,\alpha}
        \frac{M_a^b}{\sqrt{2N_c}~{\widetilde M_0}}
        ~\bar u(p_1,\lambda_1)\gamma_5\,v(p_2,\lambda_2)
                \psi_{00} (x, k_\bot)~
                |(q^c)^{a\alpha}(p_1,\lambda_1) (\bar q^c)_{b\alpha}(p_2,\lambda_2)\rangle,
        \non\\
        |M(P,\, P^{\!\!\!\!\!\!\!\!3}_0, 0)\rangle
                &=&-\int \{d^3p_1\}\{d^3p_2\} ~2(2\pi)^3 \delta^3(
                \tilde P -\tilde p_1-\tilde p_2)~\nonumber\\
        &&\times \sum_{\lambda_1,\lambda_2,\alpha}
        \frac{\widetilde
                     M_0~M_a^b}{2\sqrt{3\beta^2 N_c}~M_0}
        \bar u(p_1,\lambda_1) v(p_2,\lambda_2)
                \psi_{00} (x, k_\bot)~
                |(q^c)^{a\alpha}(p_1,\lambda_1) (\bar q^c)_{b\alpha}(p_2,\lambda_2)\rangle.
        \non\\
 \label{eq:mesonwavefunction}
\end{eqnarray}
Note that for heavy mesons with the $\bar Q q$ flavor content,
where $Q$ denotes a heavy quark, they transfer as $SU_f(3)$
triplet states. Wave functions of these states are similar to
those in the above equation, except that $q^c$ is replaced by
$Q^c$ and $M^b_a$ by $M^b$.

For the later purpose and for checking the phase convention, we
shall consider the meson decay constants.  For $J=0$ mesons, the
decay constants are defined by the matrix elements
 \be \label{eq:AM}
  \la 0|\bar q^c_2\gamma_\mu\gamma_5 q^c_1|P(P)\ra &\equiv&i  f_P P_\mu ,\qquad
      \la 0|\bar q^c_2\gamma_\mu q^c_1|S(P)\ra\equiv - f_S P_\mu,
  \en
where the $P$ and $S$ denote pseudoscalar and scalar $q_1^c \bar
q^c_2$ mesons, respectively, and an additional minus sign before
$f_S$ is due to charge conjugation. Using the relation
 \be
 \la 0|(\bar q^c)^{b'\alpha'}~\gamma^+(\gamma_5)~
 q^c_{a'\alpha'}|(q^c)^{a\alpha}(p_1,\lambda_1) (\bar
 q^c)_{b\alpha}(p_2,\lambda_2)\ra
 =\frac{N_c\delta^a_{a'}\delta^{b'}_b}{\sqrt{p_1^+p_2^+}}
 \bar v(p_2,\lambda_2)\gamma^+(\gamma_5) u(p_1,\lambda_1),
 \en
and considering $V^+$ and $A^+$ matrix elements, we
obtain~\cite{CCH}
  \be
  f_P&=&2\frac{\sqrt{2N_c}}{16\pi^3}\int dx_2 d^2 k_\bot \frac{1}{\sqrt{x_1
 x_2} \widetilde M_0}\,(m_1 x_2+m_2
 x_1)\,\varphi(x_2,k_\perp),
 \non\\
  f_S&=&2\frac{\sqrt{2N_c}}{16\pi^3}\int dx_2 d^2k_\bot \frac{\widetilde
                     M_0}{2\sqrt{3 x_1
 x_2}M^\prime_0}\,(m_1 x_2-m_2 x_1)\,\varphi_p(x_2,k_\perp).
 \label{eq:fP,fS}
 \en
It is easy to see that for $m_1=m_2$, the scalar meson wave
function is symmetric with respect to $x_1$ and $x_2$, and hence
$f_S=0$, as it should be~\cite{Suzuki,CCH04}.

\subsubsection{Vertex functions for pentaquarks}

We adopt the Jaffe-Wilczek picture~\cite{JW} for the pentaquark
$\P_{(Q)}$ which has the quark flavor content ${\bar q}[q_1
q_2][q_3 q_4]$~$({\bar Q}[q_1 q_2][q_3 q_4])$. Vertex functions
for pentaquarks in the light-front approach is first formulated in
\cite{CCH04}. For the purpose of the calculational convenience, we
shall treat the antiquark $\bar q$ as a particle $q^c$ instead of
an antiparticle, i.e. we shall use the charge conjugated
field~\cite{CCH04}. The reason for this seemingly odd choice will
become clear in later calculations.

The scalar diquark transforms as an anti-triplet in both color and
flavor spaces. We use $\phi_{a\alpha}$, where $a$ and $\alpha$ are
flavor and color indices, respectively, to denote a diquark field.
More explicitly, in the sense of color and flavor quantum numbers,
we have
 \be
 \phi_{a\alpha}\sim\epsilon_{\alpha\beta\gamma}\epsilon_{abc}
 [q^{b\beta} q^{c\gamma}].
 \en
For example, we have
$\phi_{3\alpha}\sim\epsilon_{\alpha\beta\gamma}[u^\beta
d^\gamma]$. Note that $\phi$ is not an extrapolating field
constructed from bilinear quark fields, instead it is considered
as an effectively fundamental bosonic field to describe the
degrees of freedom of the composite diquark system. In the
Jaffe-Wilczek picute~\cite{JW} the diquark pair in the even (odd)
parity pentaquark is in a $L=1~(0)$ configuration.

In the light-front approach, the pentaquark bound state with the
total momentum $P$, spin $J=1/2$ and the orbital angular momentum
of the diquark pair $L=0,1$ can be written as~\cite{CCH04}
\begin{eqnarray}
        |\P(P,L,S_z)\rangle
                =\int &&\{d^3p_1\}\{d^3p_2\}\{d^3p_3\} ~\frac{2(2\pi)^3}{\sqrt {P^+}} \delta^3(
                \tilde P -\tilde p_1-\tilde p_2-\tilde p_3)~\nonumber\\
        &&\times \sum_{\lambda_1,\alpha,\beta,\gamma,a,b,c}
                \Psi^{S_z}_L(x_1,x_2,x_3,k_{1\bot},k_{2\bot},k_{3\bot},\lambda_1)~
                C_{\alpha\beta\gamma} (F_L)_{abc}
        \non\\
        &&\times~
             \Big|(q^c)^{a\alpha}(p_1,\lambda_1) \phi^{b\beta}(p_2) \phi^{c\beta}(p_3)\Big\rangle,
 \label{eq:lfmbspentaquark}
\end{eqnarray}
where $\alpha,\beta,\gamma$ and $a,b,c$ are color and flavor
indices, respectively, $\lambda$ denotes helicity, $p_1$, $p_2$
and $p_3$ are the on-mass-shell light-front momenta,
\begin{equation}
        \tilde p=(p^+, p_\bot)~, \quad p_\bot = (p^1, p^2)~,
                \quad p^- = {m^2+p_\bot^2\over p^+},
\end{equation}
and
\begin{eqnarray}
        &&\{d^3p\} \equiv {dp^+d^2p_\bot\over 2(2\pi)^3},
        \quad \delta^3(\tilde p)=\delta(p^+)\delta^2(p_\bot),
        \nonumber \\
        &&\Big|(q^c)(p_1,\lambda_1) \phi(p_2) \phi(p_3)\Big\rangle
        = d^\dagger_{\lambda_1}(p_1) \frac{a^\dagger(p_2) a^\dagger(p_3)}{\sqrt2}|0\rangle,\\
        &&[a(p'),a^\dagger(p)] =2(2\pi)^3~\delta^3(\tilde p'-\tilde
        p),\,
        \{d_{\lambda'}(p'),d_{\lambda}^\dagger(p)\} =
        2(2\pi)^3~\delta^3(\tilde p'-\tilde p)~\delta_{\lambda'\lambda}.
                \nonumber
 \label{eq:delta}
\end{eqnarray}
The coefficient
$C_{\alpha\beta\gamma}=\epsilon_{\alpha\beta\gamma}/\sqrt6$ is a
normalized color factor and $(F_L)_{abc}$ is a normalized flavor
coefficient obeying the relation
 \be
 &&C^{*\alpha'\beta'\gamma'} (F^*_L)^{a'b'c'}
 C_{\alpha\beta\gamma} (F_L)_{abc}
         \Big \la (q^{\prime c})_{a'\alpha'} (p'_1,\lambda'_1)
                   \phi_{b'\beta'}(p'_2) \phi_{c'\gamma'}(p'_3)
             \Big|(q^c)^{a\alpha}(p_1,\lambda_1) \phi^{b\beta}(p_2)\phi^{c\gamma}(p_3)
             \Big\rangle
 \non\\
&&=2^3(2\pi)^9~\delta^3(\tilde p'_1-\tilde p_1)\frac{1}{2}[
 \delta^3(\tilde p'_2-\tilde p_2)\delta^3(\tilde p'_3-\tilde p_3)
 +(-)^L\delta^3(\tilde p'_2-\tilde p_3)\delta^3(\tilde p'_3-\tilde
 p_2)]\delta_{\lambda'_1\lambda_1}.
 \label{eq:norm}
 \en
Note that $(F_L)_{abc}$ is (anti-)symmetric under
$b\leftrightarrow c$ for $L=1~(0)$. For example, $(F_1)_{333}=1$
is the only non-vanishing element of $(F_1)_{abc}$ in the
$\Theta^+$ case and further examples are given in the Appendix. As
we shall see below, the factor of $(-)^L$ will be compensated by
the corresponding wave function under the $p_2\leftrightarrow p_3$
interchange.

In terms of the light-front relative momentum variables $(x_i,
k_{i\bot})$ for $i=1,2,3$ defined by
\begin{eqnarray}
        && p^+_i=x_i P^{+}, \quad \sum_{i=1}^3 x_i=1, \nonumber \\
        && p_{i\bot}=x_i P_\bot+k_{i\bot}, \quad \sum_{i=1}^3 k_{i\bot}=0,
\end{eqnarray}
the momentum-space wave-function $\Psi^{S_z}_L$ can be expressed
as
\begin{equation}
        \Psi^{S_z}_L(x_i,k_{i\bot},\lambda_1)
                = \langle \lambda_1|{\cal R}_M^\dagger(x_1,k_{1\bot}, m_1)|s_1\rangle~
                \la L \frac{1}{2}; m s_1|L \frac{1}{2};\frac{1}{2} S_z\ra
                  ~\Phi_{Lm}(x_1,x_2,x_3,k_{1\bot},k_{2\bot},k_{3\bot}),
\label{eq:Psi}
\end{equation}
where $\Phi_{Lm}(x_1,x_2,x_3,k_{1\bot},k_{2\bot},k_{3\bot})$
describes the momentum distribution of the constituents in the
bound state with the subsystem consisting of the particles 2 and 3
in the orbital angular momentum $L,\,L_z=m$ state, $\la
L\frac{1}{2}; m s_1|1 \frac{1}{2};\frac{1}{2} S_z\ra$ is the
corresponding Clebsch-Gordan coefficient and $\langle
\lambda_1|{\cal R}_M^\dagger(x_1,k_{1\bot}, m_1)|s_1\rangle$ is
the well normalized Melosh transform matrix element. Its explicit
form is given in Eq.~(\ref{eq:Melosh}). Note that internal
variables in this case are defined as
 \be
 M_0^2&=&\sum_{i=1}^3\frac{m_i^2+k^2_{i\bot}}{x_i},\quad
 k_i=(\frac{m_i^2+k^2_{i\bot}}{x_i M_0},x_i M_0,\,
 k_{i\bot})=(e_i-k_{iz},e_i+k_{iz},k_{i\bot}),
 \non\\
 M_0&=&e_1+e_2+e_3,\quad
 e_i =\sqrt{m^{2}_i+k^{2}_{i\bot}+k^{2}_{iz}}=\frac{x_i M_0}{2}+\frac{m_i^2+k^{2}_{i\bot}}{2 x_i M_0},\quad
 k_{iz}=\frac{x_i M_0}{2}-\frac{m_i^2+k^{2}_{i\bot}}{2 x_i M_0}.
 \non\\
 \label{eq:internal5q}
 \en
Although the same notation is applied to both meson and pentaquark
internal quantities, one should be aware of their differences~[cf.
Eqs.~(\ref{eq:M02q}), (\ref{eq:internal2q}) and
(\ref{eq:internal5q})].

In practice it is more convenient to use the covariant form for
the Melosh transform matrix element~\cite{CCH04}
\begin{equation}
       \langle \lambda_1|{\cal R}_M^\dagger(x_1,k_{1\bot}, m_1)|s_1\rangle~
                \la L \frac{1}{2}; m s_1|L \frac{1}{2};\frac{1}{2} S_z\ra
       \non\\=\frac{1}{\sqrt{2(p_1\cdot\bar P+m_1 M_0)}}
        ~\bar u(p_1,\lambda_1)\Gamma_{Lm} u(\bar P,S_z),
        \label{eq:covariant}
\end{equation}
with
\begin{eqnarray}
        &&\Gamma_{00}=1,\qquad
        \Gamma_{1m}=-\frac{1}{\sqrt3}\gamma_5\not\!\varepsilon^*(\bar
        P,m),\non\\
        &&\bar P\equiv p_1+p_2+p_3, \non \\
        &&\varepsilon^\mu(\bar P,\pm 1) =
                \left[{2\over P^+} \vec \varepsilon_\bot (\pm 1) \cdot
                \vec P_\bot,\,0,\,\vec \varepsilon_\bot (\pm 1)\right],
                \quad \vec \varepsilon_\bot
                (\pm 1)=\mp(1,\pm i)/\sqrt{2}, \nonumber\\
        &&\varepsilon^\mu(\bar P,0)={1\over M_0}\left({-M_0^2+P_\bot^2\over
                P^+},P^+,P_\bot\right),   \label{polcom}
\end{eqnarray}
for pentaquark states with $L=0$ or $L=1$ diquark pairs. It should
be remarked that in the conventional LF approach $\bar
P=p_1+p_2+p_3$ is not equal to the baryon's four-momentum  as all
constituents are on-shell and consequently $u(\bar P,S_z)$ is not
equal to $u(P,S_z)$; they satisfy different equations of motions
$(\not\!\!\bar P-M_0)u(\bar P,S_z)=0$ and $(\not\!\!P-M)u(
P,S_z)=0$. This is similar to the case of a vector meson bound
state where the polarization vectors $\varepsilon(\bar P,S_z)$ and
$\varepsilon(P,S_z)$ are different and satisfy different equations
$\varepsilon(\bar P,S_z)\cdot\bar P=0$ and
$\varepsilon(P,S_z)\cdot P=0$~\cite{Jaus91}. Although $u(\bar
P,S_z)$ is different than $u(P,S_z)$, they satisfy the relation
 \be
 \gamma^+ u(\bar P,S_z)=\gamma^+ u(P, S_z),
 \label{eq:gamma+u}
  \en
followed from $\gamma^+\gamma^+=0$, $\bar P^+=P^+$, $\bar
P_\bot=P_\bot$. This is again in analogy with the case of
$\varepsilon(\bar P,\pm 1)=\varepsilon(P,\pm 1)$. The above
relation is useful in extracting transition form factors to be
discussed later.

The pentaquark baryon state is normalized as
\begin{equation}
        \langle \P(P',S'_z)|\P(P,S_z)\rangle = 2(2\pi)^3 P^+
        \delta^3(\tilde P'- \tilde P)\delta_{L'L}\delta_{S'_z S_z}~,
\label{wavenor}
\end{equation}
so that [cf. Eqs. (\ref{eq:lfmbspentaquark}), (\ref{eq:norm}) and
(\ref{eq:Psi})]
\begin{equation}
        \int \left(\Pi_{i=1}^3{dx_i\,d^2k_{i\bot}\over 2(2\pi)^3}\right)
        2(2\pi)^3\delta(1-\sum x_i)\delta^2(\sum k_{i\bot})
        ~\Phi^*_{L'm'}(\{x\},\{k_\bot\}) \Phi_{Lm}(\{x\},\{k_\bot\})
        =~\delta_{L'L}\delta_{m'm}.
\label{momnor}
\end{equation}
Under the constraint of $1-\sum_{i=1}^3 x_i=\sum_{i=1}^3
(k_i)_{x,y,z}=0$, we have the expressions
 \be
   &&\Phi_{Lm}(\{x\},\{k_\bot\})=\sqrt{\frac{\partial(k_{2z}, k_{3z})}{\partial
   (x_2,
  x_3)}}\,\varphi_{00}(\vec k_1,\beta_1)~\varphi_{Lm}
  \left(\frac{\vec k_2-\vec k_3}{2},\beta_{23}\right), \non\\
  &&\frac{\partial(k_{2z}, k_{3z})}{\partial (x_2,
  x_3)}=\frac{e_1 e_2 e_3}{x_1 x_2 x_3 M_0},\quad
  \varphi_{00}(\vec k,\beta)=\varphi(\vec k,\beta),\quad
  \varphi_{1m}(\vec k,\beta)=k_{m} \varphi_p(\vec k,\beta),
  \label{eq:phi}
 \en
where $k_m=-\vec\varepsilon(m)\cdot\vec k=\varepsilon(\bar
P,m)\cdot p$, or explicitly $k_{m=\pm1}=\pm(k_{\bot x}\pm i
k_{\bot y})/\sqrt2$, $k_{m=0}=-k_{z}$, are proportional to the
spherical harmonics $Y_{1m}$ in momentum space, and $\varphi$,
$\varphi_p$ are the distribution amplitudes of $s$-wave and
$p$-wave states, respectively. For a Gaussian-like wave function,
one has Eq.~(\ref{eq:wavefn}) and~\cite{CCH04}
 \be
 \Phi_{1m}(\{x\},\{k_\bot\})=\sqrt{\frac{2}{\beta_{23}^2}}
 \frac{(k_2-k_3)_m}{2}\Phi_{00}(\{x\},\{k_\bot\}).
 \label{eq:phip}
 \en
By virtue of Eq.~(\ref{eq:covariant}) it is straightforward to
obtain
 \be
 &&\langle \lambda_1|{\cal R}_M^\dagger(x_1,k_{1\bot}, m_1)|s_1\rangle~
                \la 1 \frac{1}{2}; m s_1|1 \frac{1}{2};\frac{1}{2} S_z\ra \frac{(k_2-k_3)_m}{2}
 \non\\
 &&=\frac{1}{2\sqrt{6(p_1\cdot\bar P+m_1 M_0)}}
        ~\bar u(p_1,\lambda_1)\gamma_5\left[\not\!p_2-\not\!p_3-\frac{\bar P\cdot (p_2-p_3)}{M_0}\right]u(\bar
        P,S_z).
         \label{eq:covariantP}
 \en
where the factor of $(k_2-k_3)_m=\varepsilon(\bar
P,m)\cdot(p_2-p_3)$ comes from the wave function
Eq.~(\ref{eq:phip}) for the $L=1$ case. The state $|\P(P,L,
S_z)\rangle$ for a pentaquark $\P$ in the light-front model can
now be obtained by using
Eqs.~(\ref{eq:lfmbspentaquark})--(\ref{eq:covariantP}).

Putting everything together we have
\begin{eqnarray}
        |\P(P,L=0,S_z)\rangle
                =\int &&\{d^3p_1\}\{d^3p_2\}\{d^3p_3\} ~\frac{2(2\pi)^3}{\sqrt {P^+}} \delta^3(
                \tilde P -\tilde p_1-\tilde p_2-\tilde p_3)~\nonumber\\
        &&\times \sum_{\lambda_1,\alpha,\beta,\gamma,a,b,c}
                 \frac{\Phi_{00}(\{x\},\{k_\bot\})}{\sqrt{2(p_1\cdot\bar P+m_1 M_0)}}
        ~\bar u(p_1,\lambda_1) u(\bar P,S_z)
        \non\\
        &&\times \quad
             C_{\alpha\beta\gamma} (F_{L=0})_{abc}~
             \Big|(q^c)^{a\alpha}(p_1,\lambda_1) \phi^{b\beta}(p_2) \phi^{c\beta}(p_3)\Big\rangle,
 \label{eq:pentaquarkwavefunction0}
 \end{eqnarray}
for pentaquark states with $L=0$ diquark pairs, and
\begin{eqnarray}
        |\P(P,L=1,S_z)\rangle
                =\int &&\{d^3p_1\}\{d^3p_2\}\{d^3p_3\} ~\frac{2(2\pi)^3}{\sqrt {P^+}} \delta^3(
                \tilde P -\tilde p_1-\tilde p_2-\tilde p_3)~\nonumber\\
        &&\times \sum_{\lambda_1,\alpha,\beta,\gamma,a,b,c}
                \frac{\Phi_{00}(\{x\},\{k_\bot\})}{\sqrt{12~\beta_{23}^2(p_1\cdot\bar P+m_1 M_0)}}
        \non\\
        &&\times ~\bar u(p_1,\lambda_1)\gamma_5\left[\not\!p_2-\not\!p_3-\frac{\bar P\cdot (p_2-p_3)}{M_0}\right]u(\bar
        P,S_z)
        \non\\
        &&\times~ C_{\alpha\beta\gamma} (F_{L=1})_{abc}~
             \Big|(q^c)^{a\alpha}(p_1,\lambda_1) \phi^{b\beta}(p_2) \phi^{c\beta}(p_3)\Big\rangle,
 \label{eq:pentaquarkwavefunction1}
 \end{eqnarray}
for pentaquark states with $L=1$ diquark pairs. Note that these
vertex functions have been used to obtain weak transition form
factors, which are consistent with the heavy quark
symmetry~\cite{CCH04}.

\subsection{Pentaquarks strong decays}

\begin{figure}[t!]

\centerline{
            {\epsfxsize3.5 in \epsffile{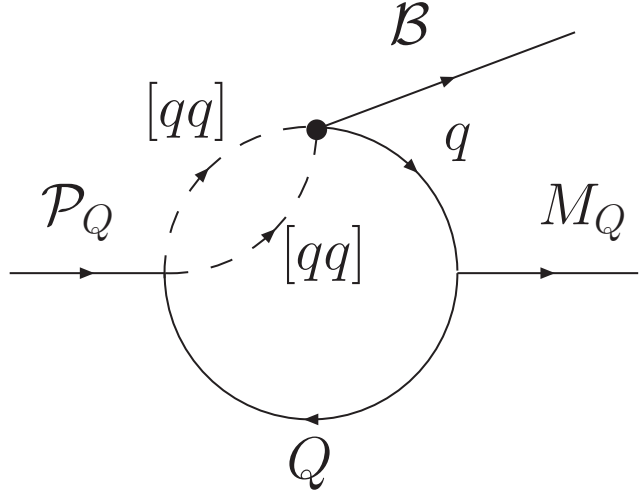}}}
\caption{Feynman diagram for a typical $\P_Q\to M_{Q}{\cal B}$
transition with $\B$ being an octet baryon, where the spin-zero
diquarks $([qq]=[ud],[us],[ds])$ are denoted by dashed lines and
the corresponding operator ${\cal O}_{\rm eff}$ (modelling the
diquark pair to the ${\cal B}q$ transition as discussed in Sec. II
B) by $\bullet$.} \label{fig:penta} 
\end{figure}

\subsubsection{Spectator approximation and the modelling of the $\phi\phi\to {\cal B} q$ sub-process}

In a typical pentaquark decay to a meson and a baryon, the
anti-quark is common to both pentaquark and the final state meson.
To the leading order of the spectator approximation, the
anti-quark can be considered as a spectator in the decay process
depicted in Fig.~1. In this picture, there is a  $\phi\phi\to
{\cal B} q$ subprocess with $\phi\phi$ being a diquark pair and
${\cal B}$ a baryon. We use the effective Hamiltonian
\footnote{Note that the $\gamma_5$ term is needed owing to the
parity conservation for strong interactions and the presence of
the $q^c$ field.}
 \be \label{eq:Heff}
 H_{\rm eff}&=&\frac{g_{1\rm eff}}{M} \epsilon^{\alpha\beta\gamma} \epsilon^{abc}
 ~\ov {\cal B}^d_c \gamma_5(q^c)_{a\alpha}
 \phi_{b\beta}\phi_{d\gamma}
 \non\\
 &+&\frac{g_{2\rm eff}}{M^2} \epsilon^{\alpha\beta\gamma} \epsilon^{abc}
 ~\ov {\cal B}^d_c i\gamma^\mu\gamma_5(q^c)_{a\alpha}
 \phi_{b\beta}\partial_\mu\phi_{d\gamma}
 \en
to model (or mimic) the $\phi\phi\to {\cal B} q$ subprocess, where
$M={\cal O}(m_\phi, m_{\cal B})$ is a characteristic scale of the
system. In general, the coupling constants $g_{1,2\rm eff}$ could
have momentum dependence. Since we are considering a soft process,
we may regard $g_{1,2\rm eff}$ as averaged and effective coupling
constants. Because the constituent quarks in octet baryons are in
the $s$-wave configuration, it is necessary to bring the two
diquarks in the pentaquark close together for interactions to
induce a strong decay. Therefore, it is plausible to use local
operators to approximate the effective Hamiltonian.

The strong decay amplitude of a pentaquark can be approximated by
 \be
 M(\P\to {\cal B}M)&\approx&\la {\cal B} M|H_{\rm eff}|\P\ra
 \non\\
                  &=&\la {\cal B}|~\bar {\cal B}^d_c|0\ra
                   \frac{g_{2\rm eff}}{M^2} \epsilon^{\alpha\beta\gamma} \epsilon^{abc}
                   \la M|i\gamma^\mu\gamma_5(q^c)_{a\alpha}
                   \phi_{b\beta}\partial_\mu\phi_{d\gamma}|\P\ra.
 \label{eq:Mstrong}
 \en
As we shall see, only the $g_{2\rm eff}$ term in Eq.
(\ref{eq:Heff}) is relevant for the strong decays of even-parity
pentaquarks under the spectator approximation. The antiquark
common to the pentaquark and to the final state meson behaves as a
spectator. Eq.~(\ref{eq:Mstrong}) can be considered as an ansatz.
In this work we shall estimate the matrix element $\la M|(q^c)
\phi\partial\phi|\P\ra$  using the light-front approach. Once the
coupling constant $g_{2\rm eff}$ is extracted from the process
such as $\Theta^+\to KN$, we can apply it to estimate other $\P\to
M{\cal B}$ strong decays.

In Eq.~(\ref{eq:Mstrong}) we have\footnote{Note that $|{\cal
B}\ra$ is normalized in the same way as Eq.~(\ref{wavenor}) and is
different from the $|q\ra$ normalization.} $\la {\cal B}|~\bar
{\cal B}^d_c|0\ra=\bar u(P_{\cal B},S'_z) T^d_c$, where $T^d_c$ is
a  traceless $3\times3$ matrix element corresponding to the
emitted baryon SU(3) quantum number and its explicit form is given
in the Appendix.  Defining
 \be
 {\cal O}_{\rm eff}=i\epsilon^{\alpha\beta\gamma} \epsilon^{abc}
                   \gamma^\mu\gamma_5(q^c)_{a\alpha}
                   \phi_{b\beta}\partial_\mu\phi_{d\gamma}T^d_c.
 \en
we can recast Eq.~(\ref{eq:Mstrong}) to
 \be
 M(\P\to M{\cal B})=\frac{g_{2\rm eff}}{M^2} \bar u(P_{\cal B},S'_z)
                   \la M|{\cal O}_{\rm eff}|\P\ra.
 \label{}
 \en
From Lorentz covariance and SU(3) symmetry, we have the general
expressions
 \be
 \la P({\bf8})|{\cal O}_{\rm eff}|\P({\bf\overline {10}})\ra
                   &=&\epsilon^{ijk} \P_{imn}T_j^m P_k^n
                   ~ f(q^2)\, i\gamma_5 u(P_\P,S_z),
                   \non\\
 \la P_Q({\bf3})|{\cal O}_{\rm eff}|\P_{Q}({\bf\bar 6})\ra
                   &=&\epsilon^{ijk} (\P_Q)_{im} T_j^m (P_Q)_k
                   ~ f_Q(q^2)\, i\gamma_5 u(P_\P,S_z),
                   \non\\
 \la S_Q({\bf3})|{\cal O}_{\rm eff}|\P_{Q}({\bf\bar 6})\ra
                   &=&-\epsilon^{ijk} (\P_Q)_{im} T_j^m (S_Q)_k
                   ~ g_Q(q^2)u(P_\P,S_z),
 \label{eq:fg}
 \en
where $f,f_Q,g_Q$ are form factors with dimension 2,
$\P({\bf\overline {10}})$ is an anti-decuplet pentaquark,
$\P_Q({\bf\bar 6})$ is a heavy anti-sextet pentaquark, $P({\bf8})$
is an octet pseudoscalar meson, $P_Q({\bf 3})$ is a heavy triplet
pseudoscalar meson and $S_Q({\bf 3})$ is a heavy triplet scalar
meson. Note that ${\cal P}_{Q}$ ($P_{Q}, S_Q$) in right hand side
of the above equation is a $3\times3$ ($3\times 1$) matrix
characterizing the SU(3) quantum numbers of the corresponding
states; that is, $\P_{ijk}=(F_{L=1})_{ijk},
(\P_Q)_{ij}=(F_{L=1})_{ij},P^l_m=M^l_m,(P_Q)_l=M_l,(S_Q)_l=M_l$.
Our approach is consistent with the generic SU(3)
approach~\cite{He}. Armed with the meson and pentaquark
(phenomenological) wave functions
[cf.~Eqs.~(\ref{eq:mesonwavefunction}),
(\ref{eq:pentaquarkwavefunction0}) and
(\ref{eq:pentaquarkwavefunction1}) ], we are ready to estimate
these form factors. To the end, we will gain more information than
that based solely on flavor symmetry. For example, it will be
interesting to see how the transition matrix elements involving
different final state mesons, such as $s$-wave and $p$-wave ones,
behave.

It is interesting to note that in the soft meson limit, the
pentaquark decay amplitude can be related to the axial-vector
matrix element $\la \B|A_\mu|\P\ra$. According to the action of
the axial current there are two possible diagrams: an annihilation
diagram and a transition one. The annihilation diagram has been
considered in~\cite{Melikhov} and it is close to the one
considered here as depicted in Fig.~\ref{fig:penta}, while the
transition diagram is the analogue of the so-called $Z$-graph. In
the present framework, the $Z$-diagram is obtained by replacing
the $\phi\phi\to{\cal B}q$ sub-process in Fig.~\ref{fig:penta} by
the $\phi\phi q^c\to{\cal B}$ one. As in
\cite{CCH,Schlumpf,CCH04}, we consider the $q^+=0$, $q_\bot\not=0$
case where the $Z$-diagram contribution is
absent~\cite{Jaus91,Cheng97}.

We shall follow~\cite{Schlumpf,CCH04} to project out various form
factors from the transition matrix elements. To extract form
factors $f,f_Q,g_Q$, we apply the relation~\cite{Pauli}
 \be
 &&\frac{\bar u(P',S'_z)\gamma^+ u(P, S_z)}{2\sqrt{P^+ P^{\prime
 +}}}\non\\
 &&=\frac{\bar u(\bar P',S'_z)\gamma^+ u(\bar P, S_z)}{2\sqrt{P^+ P^{\prime
 +}}}=\frac{u(\bar P',S'_z)\gamma^+ u(P, S_z)}{2\sqrt{P^+ P^{\prime
 +}}}=\frac{\bar u(P',S'_z)\gamma^+ u(\bar P, S_z)}{2\sqrt{P^+ P^{\prime
 +}}}=\delta_{S'_z S_z},
 \label{eq:spinorprojection}
  \en
which can be obtained by applying Eqs.~(\ref{eq:u}) and
(\ref{eq:gamma+u}), and multiplying $\bar
u(P,S_z)\gamma^+\gamma_5(=\bar u(\bar P,S_z)\gamma^+\gamma_5)$ to
the first two equations of Eq.~(\ref{eq:fg}) from the left and
$\bar u(P,S_z)\gamma^+(=\bar u(\bar P,S_z)\gamma^+)$ to the last
equation of Eq.~(\ref{eq:fg}) from the left.

\subsubsection{Even-parity Pentaquark to pseudoscalar and to scalar meson
transitions}

It is easy to derive the relation
 \be
 &&\la (q^c)_{a'\alpha'}(p'_1,\lambda'_1)(\bar q^c)^{b'\alpha'}(p'_2,\lambda'_2)|
 \gamma_5(q^c)_{a''\alpha''}~\phi_{b\beta''}i\partial_\mu\phi_{d''\gamma''}
 |(q^c)^{a\alpha}(p_1,\lambda_1)\phi^{b\beta}(p_2)\phi^{c\gamma}(p_3)\ra
 \non\\
 &&=\frac{2(2\pi)^3}{\sqrt{2p^{\prime+}_2 p^+_2 p^+_3}}~\gamma_5 v(p'_2,\lambda'_2)
          ~\delta^3(\tilde p'_1-\tilde p_1)
    ~\delta_{\lambda_1\lambda'_1}\delta_{a''}^{b'}\delta_{\alpha''}^{\alpha}\delta_{a'}^a
    [p_{3\mu}\delta_{b''}^{b}\delta_{\beta''}^{\beta}\delta_{d''}^{c}\delta_{\gamma''}^{\gamma}
    +p_{2\mu}\delta_{d''}^{b}\delta_{\gamma''}^{\beta}\delta_{b''}^{c}\delta_{\beta''}^{\gamma}],
 \non\\
 \en
where the $1/\sqrt{2}$ factor is ascribed to the identical
particles of $|\phi\phi\ra$ and is included in the initial state
as defined in Eq.~(\ref{eq:delta}). Since we do not have an
$SU_f(3)$ singlet meson in the final state, a disconnected term,
which occurs from the contraction of the final state
quark-antiquark pair, is dropped from the above equation. For the
even-parity pentaquark decay matrix element, the terms
$\epsilon^{a''b''c''}\epsilon^{\alpha''\beta''\gamma''}
T^{d''}_{c''}$ from ${\cal O}_{\rm eff}$ and
$C_{\alpha\beta\gamma}(F_{L=1})_{abc}$ from $|\P\ra$ will be
contracted with the above equation. Since $(F_{L=1})_{abc}$ and
$\epsilon^{\alpha\beta\gamma}$ are symmetric and anti-symmetric,
respectively, in interchanging any of the two indices, we are led
to a factor of $C_{\alpha\beta\gamma} \epsilon^{\alpha\beta\gamma}
(F_{L=1})_{a'b''d''}\epsilon^{b'b''c''}T^{d''}_{c''}(p_3-p_2)_\mu$
after contraction. It can be easily seen that the matrix element
will be vanished if $\phi\phi$ rather than $\phi\partial\phi$ is
employed in ${\cal O}_{\rm eff}$. This is the reason why only the
$g_{2\rm eff}$ term in $H_{\rm eff}$ contributes.

For $\P({\bf\overline{10}})\to P({\bf 8})$ transitions, we have
 \be
 \la P(P')|{\cal O}_{\rm eff}|\P(P,S_z)\ra
 &=&-i\int\{d^3 p_1\}\{d^3 p_2\}
 \frac{\epsilon^{lij} (F_{L=1})_{min} T^n_{j} M^m_l
 }{2\sqrt{2~\beta_{23}^2(p_1\cdot\bar P+m_1
  M_0) P^+ p^{\prime+}_2 p^+_2 p^+_3 N_c}\widetilde
  M'_0}
  \non\\
        &\times&(\not\!p_2-\not\!p_3)\gamma_5 (\not\!p'_2-m'_2)\gamma_5 (\not\!p_1+m_1)
        \gamma_5\left[\not\!p_2-\not\!p_3-\frac{\bar P\cdot (p_2-p_3)}{M_0}\right]
  \non\\
  &\times&  u(\bar P,S_z)~\psi_{00}(x',k'_\bot)~\Phi_{00}(\{x\},\{k_\bot\}),
  \label{eq:P->M}
   \en
with $p'_1=p_1$, $(p_2+p_3-p'_2-q)^+=(p_2+p_3-p'_2-q)_\bot=0$,
$q^+=0$ or, equivalently,
 \be
 x'_1=x_1~({\rm or}\,\,\, x'_2=x_2+x_3),\quad
 k'_{1\bot}-x_1 q_\bot=k_{1\bot}(=-k_{2\bot}-k_{3\bot}),
 \label{eq:momentumconservation}
 \en
where $C_{\alpha\beta\gamma} \epsilon^{\alpha\beta\gamma}=\sqrt6$
and a relabelling of dummy indices has been made in
Eq.~(\ref{eq:P->M}).

Likewise, for the case of $\P_Q(\6bar)\to P_Q({\bf 3})$
transitions, we have
 \be
 \la P_Q(P')|{\cal O}_{\rm eff}|\P_Q(P,S_z)\ra
 &=&-i\int\{d^3 p_1\}\{d^3 p_2\}
 \frac{\epsilon^{lij}(F_{L=1})_{im} T^m_{j} M_l
 }
 {2\sqrt{2~\beta_{23}^2(p_1\cdot\bar P+m_1
  M_0) P^+ p^{\prime+}_2 p^+_2 p^+_3 N_c}\widetilde
  M'_0}
  \non\\
        &\times& (\not\!p_2-\not\!p_3)\gamma_5 (\not\! p'_2-m'_2)\gamma_5 (\not\! p_1+m_1)
        \gamma_5\left[\not\!p_2-\not\!p_3-\frac{\bar P\cdot (p_2-p_3)}{M_0}\right]
  \non\\
  &\times&  u(\bar P,S_z)~\psi_{00}(x',k'_\bot)~\Phi_{00}(\{x\},\{k_\bot\}),
  \label{eq:PQ->MQ}
   \en
while for the case of $\P_Q(\bar 6)\to S_Q({\bf 3})$ transitions,
 \be
 \la S_Q(P')|{\cal O}_{\rm eff}|\P_Q(P,S_z)\ra
 &=&\int\{d^3 p_1\}\{d^3 p_2\}
 \frac{\epsilon^{lij} (F_{L=1})_{im} T^m_{j} M_l\widetilde M'_0}
 {4\sqrt{3~\beta_{23}^2\beta^{\prime2}(p_1\cdot\bar P+m_1
  M_0) P^+ p^{\prime+}_2 p^+_2 p^+_3 N_c}
  M'_0}
  \non\\
        &\times& (\not\!p_2-\not\!p_3)\gamma_5 (\not\! p'_2-m'_2)(\not\! p_1+m_1)
        \gamma_5\left[\not\!p_2-\not\!p_3-\frac{\bar P\cdot (p_2-p_3)}{M_0}\right]  \non\\
  &\times&  u(\bar
  P,S_z)~\psi_{00}(x',k'_\bot)~\Phi_{00}(\{x\},\{k_\bot\}).
  \label{eq:PQ->SQ}
   \en

Multiplying $\bar u(P,S_z)\gamma^+\gamma_5$ ($\bar u(\bar
P,S_z)\gamma^+\gamma_5$) to the left (right) hand side of
Eqs.~(\ref{eq:P->M}) and (\ref{eq:PQ->MQ}) and  noting that SU(3)
factors in Eq.~(\ref{eq:fg}) and in Eqs.~(\ref{eq:P->M}),
(\ref{eq:PQ->MQ}) are the same [i.e. $\epsilon^{ijl} \P_{inm}T_j^n
P_l^m=\epsilon^{lij} (F_{L=1})_{min} T^n_{j} M^m_l,\epsilon^{ijl}
(\P_Q)_{im} T_j^m (P_Q)_l=\epsilon^{lij} (F_{L=1})_{im} T^m_{j}
M_l$] and hence can be factored out, we have
 \be
 f(q^2)
 &=&-\int\frac{dx_1 d^2k_{1\bot}}{2(2\pi)^3}\frac{dx_2 d^2k_{2\bot}}{2(2\pi)^3}
              \frac{1}{8P^+\widetilde M'_0\sqrt{2~\beta_{23}^2(p_1\cdot\bar P+m_1 M_0)  x'_2 x_2 x_3 N_c}}
  \non\\
        &\times& {\rm Tr}\Bigg\{(\not\!\bar P+M_0)\gamma^+(\not\!p_2-\not\!p_3)(\not\!p'_2-m'_2)(\not\! p_1-m_1)
         \left[\not\!p_2-\not\!p_3-\frac{\bar P\cdot
        (p_2-p_3)}{M_0}\right]\Bigg\}
   \non\\
        &\times&\psi_{00}(x',k'_\bot)~\Phi_{00}(\{x\},\{k_\bot\}),
  \non\\
  f_Q(q^2)
  &=&-\int\frac{dx_1 d^2k_{1\bot}}{2(2\pi)^3}\frac{dx_2 d^2k_{2\bot}}{2(2\pi)^3}
               \frac{1}{8P^+\widetilde M'_0\sqrt{2~\beta_{23}^2(p_1\cdot\bar P+m_1 M_0)  x'_2 x_2 x_3 N_c}
                        }
  \non\\
        &\times& {\rm Tr}\Bigg\{(\not\!\bar P+M_0)\gamma^+(\not\!p_2-\not\!p_3)(\not\! p'_2-m'_2)(\not\!\bar p_1-m_1)
              \left[\not\!p_2-\not\!p_3-\frac{\bar P\cdot (p_2-p_3)}{M_0}\right]\Bigg\}
  \non\\
        &\times&\psi_{00}(x',k'_\bot)~\Phi_{00}(\{x\},\{k_\bot\}).
    \label{eq:f}
   \en
Similarly by multiplying $\bar u(P,S_z)\gamma^+$ $(\bar u(\bar
P,S_z)\gamma^+)$ to the left (right) hand side of
Eq.~(\ref{eq:PQ->SQ}) and using the fact that $\epsilon^{ijl}
(\P_Q)_{im} T_j^m (S_Q)_l=\epsilon^{lij} (F_{L=1})_{im} T^m_{j}
M_l$, we arrive at
 \be
 g_Q(q^2)
 &=&-\int\frac{dx_1 d^2k_{1\bot}}{2(2\pi)^3}\frac{dx_2 d^2k_{2\bot}}{2(2\pi)^3}
 \frac{\widetilde M'_0}
 {16P^+M'_0\sqrt{3~\beta_{23}^2\beta^{\prime2}(p_1\cdot\bar P+m_1 M_0) x'_2 x_2 x_3 N_c}}
  \non\\
        &\times& {\rm Tr}\Bigg\{(\not\!\bar P+M_0)\gamma^+(\not\!p_2-\not\!p_3)(\not\! p'_2+m'_2)(\not\! p_1-m_1)
        \left[\not\!p_2-\not\!p_3-\frac{\bar P\cdot
        (p_2-p_3)}{M_0}\right]\Bigg\}
  \non\\
  &\times &\psi_{00}(x',k'_\bot)~\Phi_{00}(\{x\},\{k_\bot\}).
   \label{eq:g}
   \en

For a more explicit expression of above form factors we need to
work out the corresponding traces. It is straightforward to obtain
 \be
 &&\frac{1}{4P^+}{\rm Tr}\Bigg\{(\not\!\bar P+M_0)\gamma^+(\not\!p_2-\not\!p_3)(\not\!p'_2\mp m'_2)(\not\! p_1-m_1)
         \left[\not\!p_2-\not\!p_3-\frac{\bar P\cdot
        (p_2-p_3)}{M_0}\right]\Bigg\}
 \non\\
 &&\qquad=p_{23}^2[(p_1\cdot p'_2\pm m_1 m'_2)+x'_2(p_1\cdot \bar P+m_1
 M_0)-x_1(p'_2\cdot \bar P\mp m'_2M_0)]
 \non\\
 &&\qquad\quad+2x_{23} p_{23}\cdot p_1 (p'_2\cdot\bar P\mp m'_2 M_0)
              -2x_{23} p_{23}\cdot p'_2 (p_1\cdot\bar P+m_1 M_0)
 \non\\
 &&\qquad\quad+2 p_{23}\cdot \bar P (x_1 p_{23}\cdot p'_2-x'_2 p_{23}\cdot p_1)
 \non\\
 &&\qquad\quad-\frac{\bar P\cdot p_{23}}{M_0} \{p_{23}\cdot\bar P(\pm x_1
 m'_2+x'_2 m_1)+x_{23}[M_0 (p_1\cdot p'_2\pm m_1 m'_2)-m_1(p'_2\cdot
 \bar P\mp m'_2 M_0)
 \non\\
 &&\qquad\quad\mp m'_2(p_1\cdot \bar P+m_1 M_0)]-p_{23}\cdot p_1 (\pm m'_2+x'_2
 M_0)-p_{23}\cdot p'_2 (m_1-x_1M_0)\},
 \en
where use of $p_{23}\equiv p_2-p_3$, $x_{23}\equiv x_2-x_3$ has
been made.
To recast the above expression in terms of internal variables, it
is useful to note that $p_1=p'_1,\,\bar P-\bar
P'=p_2+p_3-p'_2=\bar q$, where $\bar q^{+}=q^+=0$, $\bar q_\bot=
q_\bot$, and $\bar q^2=q^2=-q_\bot^2$, $\bar q\cdot(x^{(\prime)}_i
\bar P^{(\prime)}-p_i^{(\prime)})=q_\bot\cdot
k^{(\prime)}_{i\bot}$. We then obtain
 \be
 &&\bar P\cdot \bar P'=\frac{1}{2}(M_0^2+M_0^{\prime 2}-q^2),
   \qquad
 p_1\cdot p'_2\pm m_1 m'_2=\frac{M_0^{\prime2}-(m_1\mp m'_2)^2}{2},
 \non\\
 &&p_1\cdot \bar P+m_1 M_0=\frac{(m_1+x_1 M_0)^2+k^2_{1\bot}}{2 x_1},
     \qquad
   p'_2\cdot \bar P\mp m'_2 M_0=\frac{(m'_2\mp x'_2 M_0)^2+(k_{1\bot}+q_\bot)^2}{2x'_2},
 \non\\
 &&p_{23}^2=k^2_{23}=(k^+_2-k_3^+)(k^-_{2}-k^-_{3})-k_{23\bot}^2,
     \qquad
 \bar P\cdot p_{23}=M_0 e_{23}=p_1\cdot p_{23}+m_2^2-m^2_3,
 \non\\
 &&\bar q\cdot p_{23}=-\bar q\cdot (x_{23} \bar P-p_{23})+x_{23}\bar q\cdot\bar P
                     =-q_\bot\cdot k_{23\bot}+\frac{x_{23}}{2}(M_0^2-M_0^{\prime2}+q^2),
 \non\\
 &&p'_2\cdot p_{23}=(p_2+p_3-\bar q)\cdot p_{23}
                   =\frac{2q_\bot\cdot k_{23\bot}-x_{23}(M_0^2-M_0^{\prime2}+q^2)+2(m_2^2-m^2_3)}{2},
  \label{eq:internal}
 \en
where uses of Eq.~(\ref{eq:momentumconservation}), $e_{23}\equiv
e_2-e_3$ and $k_{23}\equiv k_2-k_3$ have been made. Finally
putting these together, we obtain
 \be
f(q^2)
 &=&-\int\frac{dx_1 d^2k_{1\bot}}{2(2\pi)^3}\frac{dx_2 d^2k_{2\bot}}{2(2\pi)^3}
              \frac{1}{4 x'_2\widetilde M'_0\sqrt{\beta_{23}^2[(m_1+x_1 M_0)^2+k^2_{1\bot}]  x_1 x'_2 x_2 x_3 N_c}}
  \non\\
        &\times& \Big\{x_1 x'_2[M_0^{\prime2}-(m_1-m'_2)^2](k^2_{23} - x_{23} M_0 e_{23})
 \non\\
  &&+x'_2 [(m_1+x_1 M_0)^2+k^2_{1\bot}]
  \{[k^2_{23} x'_2+x_{23}[-2q_\bot\cdot k_{23\bot}
  \non\\
  &&+x_{23}(M_0^2-M_0^{\prime2}+q^2)-2(m_2^2-m^2_3)+m'_2
  e_{23}]\}
 \non\\
  &&+x_1 [(m'_2-x'_2 M_0)^2+(k_{1\bot}+q_\bot)^2]
  [-x_1 k^2_{23}+2 x_{23} (M_0 e_{23}-m_2^2+m_3^2)+x_{23} m_1 e_{23}]
 \non\\
  &&+x_1 x'_2 e_{23}\{-2M_0 e_{23}(x_1 m'_2 + x'_2 m_1)+ 2(M_0 e_{23}-m^2_2+m_3^2) (m'_2 - x'_2 M_0)
  \non\\
  &&+ [2q_\bot\cdot k_{23\bot}-x_{23}(M_0^2-M_0^{\prime2}+q^2)+2(m_2^2-m^2_3)]
      (m_1 + x_1 M_0)\}\Big\}
   \non\\
        &\times&\psi_{00}(x',k'_\bot)~\Phi_{00}(\{x\},\{k_\bot\}),
  \non\\
f_Q(q^2)
  &=&-\int\frac{dx_1 d^2k_{1\bot}}{2(2\pi)^3}\frac{dx_2 d^2k_{2\bot}}{2(2\pi)^3}
               \frac{1}{4 x'_2\widetilde M'_0\sqrt{\beta_{23}^2[(m_1+x_1 M_0)^2+k^2_{1\bot}] x_1 x'_2 x_2 x_3 N_c}
                        }
  \non\\
        &\times& \Big\{x_1 x'_2[M_0^{\prime2}-(m_1-m'_2)^2](k^2_{23} - x_{23} M_0 e_{23})
 \non\\
  &&+x'_2 [(m_1+x_1 M_0)^2+k^2_{1\bot}]
  \{[k^2_{23} x'_2+x_{23}[-2q_\bot\cdot k_{23\bot}
  \non\\
  &&+x_{23}(M_0^2-M_0^{\prime2}+q^2)-2(m_2^2-m^2_3)+m'_2
  e_{23}]\}
 \non\\
  &&+x_1 [(m'_2- x'_2 M_0)^2+(k_{1\bot}+q_\bot)^2]
  [-x_1 k^2_{23}+2 x_{23} (M_0 e_{23}-m_2^2+m_3^2)+x_{23} m_1 e_{23}]
 \non\\
  &&+x_1 x'_2 e_{23}\{-2M_0 e_{23}(x_1 m'_2 + x'_2 m_1)+ 2(M_0 e_{23}-m^2_2+m_3^2) (m'_2 - x'_2 M_0)
  \non\\
  &&+ [2q_\bot\cdot k_{23\bot}-x_{23}(M_0^2-M_0^{\prime2}+q^2)+2(m_2^2-m^2_3)]
      (m_1 + x_1 M_0)\}\Big\}
  \non\\
        &\times&\psi_{00}(x',k'_\bot)~\Phi_{00}(\{x\},\{k_\bot\}),
  \non\\
 g_Q(q^2)
 &=&-\int\frac{dx_1 d^2k_{1\bot}}{2(2\pi)^3}\frac{dx_2 d^2k_{2\bot}}{2(2\pi)^3}
 \frac{\widetilde M'_0}
 {4 x'_2 M'_0\sqrt{6\beta_{23}^2\beta^{\prime2}[(m_1+x_1 M_0)^2+k^2_{1\bot}] x_1 x'_2 x_2 x_3 N_c}}
  \non\\
   &\times& \Big\{x_1 x'_2[M_0^{\prime2}-(m_1+m'_2)^2](k^2_{23} - x_{23} M_0 e_{23})
 \non\\
  &&+x'_2 [(m_1+x_1 M_0)^2+k^2_{1\bot}]
  \{[k^2_{23} x'_2+x_{23}[-2q_\bot\cdot k_{23\bot}
  \non\\
  &&+x_{23}(M_0^2-M_0^{\prime2}+q^2)-2(m_2^2-m^2_3)-m'_2 e_{23}]\}
 \non\\
  &&+x_1 [(m'_2+x'_2 M_0)^2+(k_{1\bot}+q_\bot)^2] [-x_1 k^2_{23}+2 x_{23} (M_0 e_{23}-m_2^2+m_3^2)+x_{23} m_1 e_{23}]
 \non\\
  &&+x_1 x'_2 e_{23}\{2M_0 e_{23}(x_1 m'_2 - x'_2 m_1)- 2(M_0 e_{23}-m^2_2+m_3^2) (m'_2+x'_2 M_0)
  \non\\
  &&+ [2q_\bot\cdot k_{23\bot}-x_{23}(M_0^2-M_0^{\prime2}+q^2)+2(m_2^2-m^2_3)] (m_1 + x_1 M_0)\}\Big\}
  \non\\
  &\times &\psi_{00}(x',k'_\bot)~\Phi_{00}(\{x\},\{k_\bot\}).
   \label{eq:ffQgQ}
   \en
Numerical estimations of these form factors will be given in the
next section.

\section{Numerical Results}
\subsection{Strong decays of pentaquark baryons}

\begin{table}[b]
\caption{\label{tab:input} The pentaquark masses and input
parameters $m_{[qq']}$, $m_q$ and $\beta$'s (in units of GeV)
appearing in the Gaussian-type wave function (\ref{eq:wavefn}).}
\begin{ruledtabular}
\begin{tabular}{ccccccc}
            $m_{[ud]}$
          & $m_{[us]}$
          & $m_u$
          & $m_s$
          & $m_c$
          & $\beta_{1c}$
          & $\beta_{1s}$
          \\
\hline
           $0.40$
          & $0.56$
          & $0.23$
          & $0.45$
          & $1.3$
          & $0.58$
          & 0.48
          \\
\hline \hline
           $\beta_{23[qq']}$
          & $\beta_\pi$
          & $\beta_K$
          & $\beta_D$
          & $\beta_{D_s}$
          & $\beta_{D^*_{s0}}$
          \\
\hline
            0.38
          & 0.35
          & 0.377
          & 0.456
          & 0.478
          & 0.340
\end{tabular}
\end{ruledtabular}
\end{table}

The input parameters $m_{[qq']}$, $m_q$, $\beta_M$, $\beta_1$ (for
the anti-quark) and $\beta_{23}$ (for the diquark pair) [see Eq.
(\ref{eq:phi})] that are relevant for our proposes are summarized
in Table~\ref{tab:input}. The quark masses and $\beta_M$'s are
taken from \cite{CCH,CC} where the latter are obtained by fitting
to the decay constants [cf. Eq.~(\ref{eq:fP,fS})] as done in
\cite{CC}. Note that our prediction
$f_{D_{s0}^*}=59$~MeV~\cite{CCH} is consistent with the recent
experimental result $f_{D_{s0}^*}\approx 47-73~{\rm MeV}$
\cite{BelleDs,Cheng:2003id}. This supports a smaller value of
$\beta_{D_{s0}^*}$ as shown in Table~\ref{tab:input}. Since the
diquark pair acts like ${\bf 3}_c$, the $\bar q$--$\{[ud][ud]\}$
system can be regarded as the analog of the heavy meson $\bar
q$--$q'$. Therefore, it is plausible to assume that
$\beta_{1c}:\beta_{1s}\sim\beta_{D}:\beta_{K}$~\cite{CCH04}. The
$\beta_{23[qq']}$ parameter for the diquark pair is taken to be of
order $\Lambda_{\rm QCD}$. The explicit numerical values of
$\beta_{D,K}$ are taken from \cite{CCH,CC}. As shown in
\cite{CCH04}, by using these input parameters, the obtained
$\Sigma'_{5b}\to\Sigma'_{5c}$ transition form factors $f_1(0)$,
$g_1(0)$ are close to their counterparts (in the sense of
$SU_f(3)$ representation) in  the $\Lambda_b\to\Lambda_c$
transition \cite{Cheng97a}.

To proceed, we find that the momentum dependence of the form
factors in the spacelike region can be well parameterized and
reproduced in the three-parameter form:
 \be \label{eq:FFpara}
 F(q^2)&=&\,{F(0)\over
 1-(q^2/\Lambda_1^2)+(q^2/\Lambda_2^2)^2}
 \en
for $\P\to M$ transitions. The parameters $\Lambda_{1,2}$, and
$F(0)$ are first determined in the spacelike region. We then
employ this parametrization to determine the physical form factors
at $q^2\geq 0$.

Table~\ref{tab:fg} gives various form factors obtained in the
light-font approach. The from factors obtained from
Eqs.~(\ref{eq:f}), (\ref{eq:g}) are fitted to the form of
Eq.~(\ref{eq:FFpara}). For pentaquarks in which the two diquarks
have different flavors, e.g. the $\Sigma^0_{5c}(\bar c[ud][us])\to
D^-_s$ transition, we need to average over $f(q^2)$ by  applying
$m_2=m_{ud}$, $m_3=m_{us}$ followed by an interchange of them in
Eqs.~(\ref{eq:f}). We shall fit the form factors to the range of
$-3~{\rm GeV}^2<q^2\leq 0$ for light pentaquark transitions, and
to the range of $-7~{\rm GeV}^2\leq q^2\leq 0$ for heavy
pentaquark transitions. Note that for $\Theta\to K$ and
$\Xi_{3/2}\to\pi,K$ transition form factors a monopole form for
their momentum dependence is adequate; an inclusion of the
$\Lambda_2$ term will not affect the fit quality. For the case of
heavy pentaquark transitions, since the form factors are fitted to
a larger range of $q^2$, it is necessary to include the
$\Lambda_2$ term in order to achieve a better fit.

Several remarks are in order: (i) It is interesting to note from
Table~\ref{tab:fg} that $\Theta\to K$ and $\Theta_c\to D$ form
factors are very similar owing to the underlying spectator picture
in which $\bar s$ and $\bar c$ are spectators. (ii) It is
important to point out that the form factors of interest are
indeed small as one can check explicitly that $f(q^2)/m_\P^2,\,
f_Q(q^2)/m_{\P_Q}^2,\, g_Q(q^2)/m_{\P_Q}^2\ll 1$. The smallness of
form factors is ascribed to the $p$-wave configuration of the two
diquarks in an even-parity pentaquarks as it is necessary to bring
the two diquarks close together to get involved interactions and
produce an ordinary baryon with a $s$-wave quark configuration. As
noted in passing, this phenomenon has been modelled in the present
work by applying a local operator ${\cal O}_{\rm eff}$ for the
$\phi\phi\to{\cal B}q$ transition. The mismatch in the orbital
angular momentum configuration is the key physical reason for the
smallness of these form factors. (iii) The $\P_c\to D_{s0}^*$ form
factors are smaller than the $\P_c\to D_s$ ones by a factor of 2
owing to the smallness of $\beta_{D_{s0}^*}$. (iv) All the form
factors are sensitive to $\beta_{23}$, for example, $f^{\Theta\to
K}(0)=0.077$~GeV$^2$ (see Table~\ref{tab:fg}) will be enhanced by
16\% if $\beta_{23}$ is changed from 0.38~GeV to 0.42~GeV. For
pentaquark weak decays considered in \cite{CCH04}, diquarks are
spectators and hence weak decays are not sensitive to
$\beta_{23}$. In this work, diquarks are no longer spectators and
hence the strong transition form factors are sensitive to
$\beta_{23}$. However, as the pentaquark decay rates are
normalized to the $\Theta\to NK$ one, the $\beta_{23}$ dependence
will be reduced.

\begin{table}[b!]
\caption{\label{tab:fg} The transition form factors for various
pentaquark to meson transitions.}
\begin{ruledtabular}
\begin{tabular}{cccc|cccc}
           $F^{\P\to M}$
          & $F(0)$~(GeV$^2)$
          & $\Lambda_1$~(GeV)
          & $\Lambda_2$~(GeV)
          & $F^{\P\to M}$
          & $F(0)$~(GeV$^2)$
          & $\Lambda_1$~(GeV)
          & $\Lambda_2$~(GeV)
          \\
\hline     $f^{\Theta\to K}$
          & 0.077
          & 2.18
          & --
          & $f^{\Sigma_{5c}\to D_s}_c$
          & 0.045
          & 2.25
          & 2.40
          \\
          $f^{\Xi_{3/2}\to \pi}$
          & 0.065
          & 2.59
          & --
          & $g^{\Sigma_{5c}\to D_{s0}^*}_c$
          & 0.024
          & 2.64
          & 2.32
          \\
          $f^{\Xi_{3/2}\to K}$
          & 0.085
          & 2.56
          & --
          & $f^{\Xi_{5c}\to D_s}_c$
          & 0.084
          & 2.16
          & 2.55
          \\
          $f^{\Theta_{c}\to D}_c$
          & 0.081
          & 2.06
          & 2.39
          &$g^{\Xi_{5c}\to D_{s0}^*}_c$
          & 0.045
          & 2.97
          & 2.23
          \\
\end{tabular}
\end{ruledtabular}
\end{table}

With the numerical results of strong transition form factors given
in Table~\ref{tab:fg}, we are ready to estimate the corresponding
strong decays. The $\P\to{\cal B}M$ decay amplitudes are given by
 \be
 {\cal A}(\P\to{\cal B}M)&=&\bar u(P_{\cal B},S'_z)(A+i B\gamma_5)u(P_\P,S_z),
 \en
with ($\kappa\equiv g_{2\rm eff}/M^2$)
 \be
 B[\P({\bf\overline {10}})\to {\cal B}P({\bf8})]&=&\epsilon^{ijk} \P_{imn}T_j^m P_k^n
                   ~\kappa f(m^2_{\cal B}),
 \non\\
 B[\P_Q({\6bar})\to {\cal B}P_Q({\bf3})]&=&\epsilon^{ijk} (\P_Q)_{im}T_j^m (P_Q)_k
                   ~\kappa f_Q(m^2_{\cal B}),
  \non\\
 A[\P_Q({\6bar})\to {\cal B}S_Q({\bf3})]&=& -\epsilon^{ijk} (\P_Q)_{im}T_j^m (P_Q)_k
                   ~\kappa g_Q(m^2_{\cal B}),
 \non\\
 A[\P({\bf\overline {10}})\to {\cal B}P({\bf8})]
 &=&A[\P_Q({\6bar})\to {\cal B}P_Q({\bf3})]
 =B[\P_Q({\6bar})\to {\cal B}P_Q({\bf3})]=0,
 \label{eq:amplitude}
 \en
followed from Eqs.~(\ref{eq:Mstrong}) and (\ref{eq:fg}). The
explicit expression of the Clebsch-Gordan coefficients in
Eq.~(\ref{eq:amplitude}) can be found in the Appendix. For the
decay modes under consideration, the non-vanishing amplitudes read
 \be
 B(\Theta\to pK^0)&=&-B(\Theta\to nK^+)=\kappa f^{\Theta\to K}(m^2_N),
 \non\\
 B(\Xi^{--}_{3/2}\to \Xi^-\pi^-)&=&\kappa f^{\Xi_{3/2}\to \pi}(m^2_{\Xi}),
 \non\\
 B(\Xi^{--}_{3/2}\to \Sigma^-K^-)&=&\kappa f^{\Xi_{3/2}\to K}(m^2_{\Sigma}),
 \non\\
 B(\Theta^0_c\to p D^-)&=&\kappa f^{\Theta_c\to D}_c(m^2_p),
  \non\\
 B(\Sigma^0_{5c}\to p D^-_s)&=&-\frac{\kappa}{\sqrt{2}}~f^{\Sigma_{5c}\to D_s}_c(m^2_p),
  \non\\
 A(\Sigma^0_{5c}\to p D^{*-}_{s0})&=&\frac{\kappa}{\sqrt{2}}~g^{\Sigma_{5c}\to D^*_{s0}}_c(m^2_p),
  \non\\
 B(\Xi^0_{5c}\to \Sigma^+ D^-_s)&=&-\kappa f^{\Xi_{5c}\to D_s}_c(m^2_\Sigma),
  \non\\
 A(\Xi^0_{5c}\to \Sigma^+ D^{*-}_{s0})&=&\kappa g^{\Xi_{5c}\to D^*_{s0}}_c(m^2_\Sigma),
 \label{eq:amplitudeexplicit}
 \en
where the quark flavor content of the sextet charmed pentaquarks
is explained in \cite{CCH04}.
The decay rate can be evaluated via \cite{Cheng97a}
 \be
 \Gamma(\P\to
 {\cal B}M)&=&\frac{p_c}{8\pi}\left[\frac{(m_\P+m_{\cal B})^2-m_{M}^2}{m_\P^2}|A|^2
                                   +\frac{(m_\P-m_{\cal B})^2-m_{M}^2}{m_\P^2}|B|^2\right],
 \label{eq:rate}
 \en
where $p_c$ is the c.m. momentum of the final state in the
pentaquark rest frame.

By fitting to $\Gamma(\Theta\to
pK^0)=\frac{1}{2}\Gamma(\Theta)\simeq 0.5$~MeV, we obtain $\kappa
f(m_N^2)\simeq 0.97$. Using the result of $f(m_N^2)=0.095$~GeV$^2$
from Table~\ref{tab:fg}, it follows that $\kappa\equiv g_{2\rm
eff}/M^2\simeq 10.2$~GeV$^{-2}$, where $M$ is a characteristic
scale of the $\phi\phi\to{\cal B}q$ transition. Taking $M\simeq
1$~GeV, we have $g_{2\rm eff}\simeq 10.2$, which is slightly
smaller than the strong $\pi NN$ coupling $g_{\pi NN}\sim 14$.
This suppression could be understood as the cost to pay for
breaking one of the diquarks into two quarks.

\begin{figure}[t!]

\centerline{
            {\epsfxsize3 in \epsffile{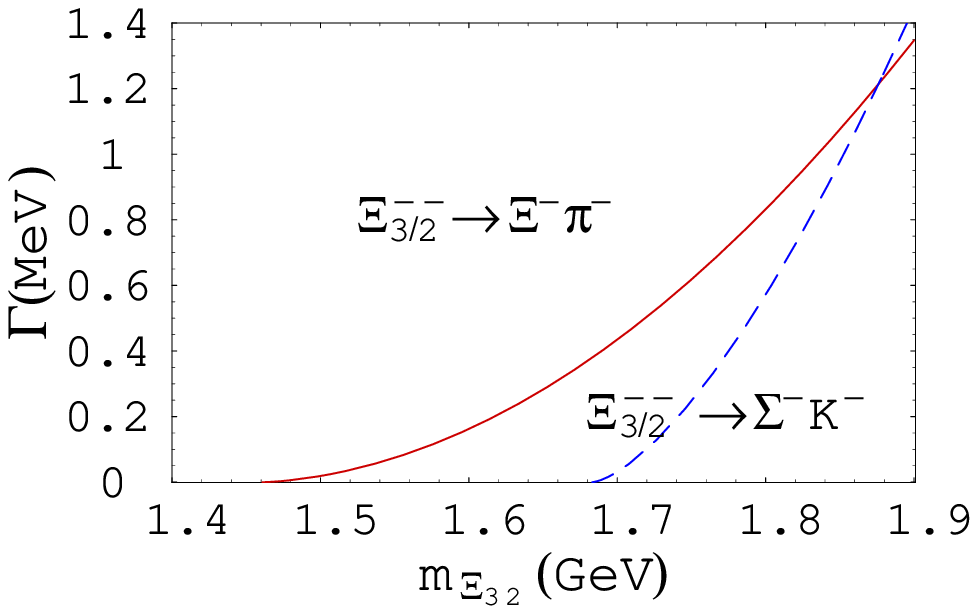}}{\epsfxsize3 in \epsffile{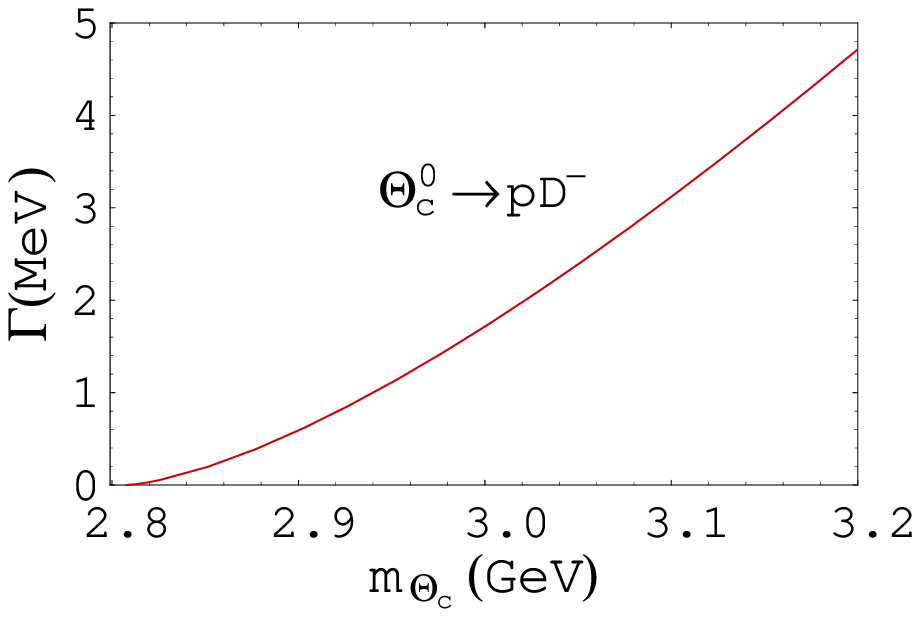}}
            }
\smallskip
\centerline{
            {\epsfxsize3 in \epsffile{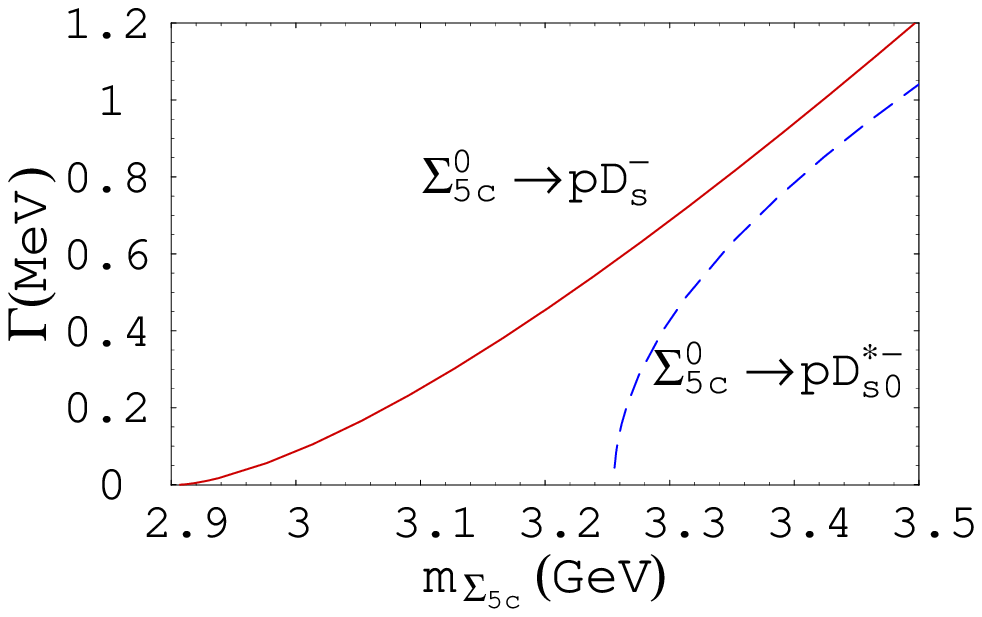}}{\epsfxsize3 in \epsffile{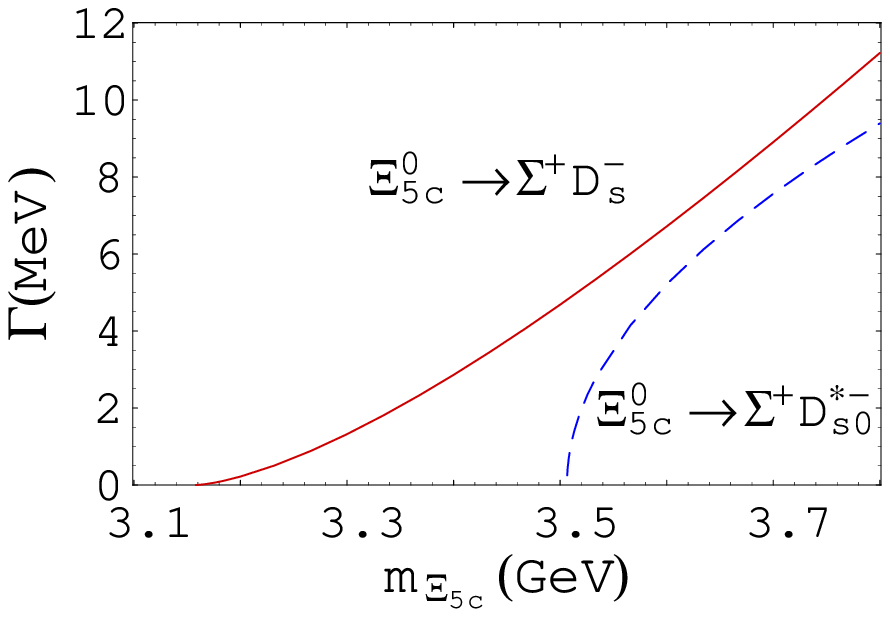}}
            }
\caption{Decay rates (in units of MeV) for
$\Xi^{--}_{3/2}\to\Xi^-\pi^-,\Sigma^- K^-$, $\Theta^0_c\to pD^-$,
$\Sigma_{5c}\to p D^-_s, pD^{*-}_{s0}$ and $\Xi^0_{5c}\to \Sigma^+
D^-_s,\Sigma^+D^{*-}_{s0}$ as a function of the pentaquark mass.
These rates are normalized to
$\Gamma(\Theta^+)=2\Gamma(\Theta^+\to pK^0)=1$~MeV.}
\label{fig:rate}
\end{figure}

With the effective strong coupling in a reasonable size, it is now
plausible to ascribe the narrow width of $\Theta^+$  to the
suppressed transition form factors ($f^{\Theta\to
K}/m_{\Theta}^2\ll 1$). As noted in passing, this suppression
arises from bringing the two diquarks in a $p$-wave configuration
close together to form a final state baryon in the $s$-wave quark
configuration.

Treating $\kappa$ to be approximately universal, we can estimate
the strong decay rates of $\Xi^{--}_{3/2}\to\Xi^-\pi^-,\Sigma^-
K^-$, $\Theta^0_c\to pD^-$, $\Sigma^0_{5c}\to p D^-_s,
pD^{*-}_{s0}$ and $\Xi^0_{5c}\to \Sigma^+
D^-_s,\Sigma^+D^{*-}_{s0}$.  In Fig.~\ref{fig:rate}, we show these
rates normalized to $\Gamma(\Theta)=2\Gamma(\Theta\to pK^0)=1$~MeV
as a function of the pentaquark mass~\footnote{In this estimation
the dependence of the pentaquark mass in rates are explicitly
shown in Eq.~(\ref{eq:rate}) with $A$ and $B$ terms being kept
fixed in the mass range under consideration.}. It is clear that
$\Xi^{--}_{3/2}\to\Xi^-\pi^-,\Sigma^-K^-$, $\Sigma^0_{5c}\to p
D^-_s, pD^{*-}_{s0}$ and $\Theta^0_c\to pD^-$ decay rates are of
order a few MeV, while $\Xi_{5c}^0\to \Sigma^+ D^-_s,
\Sigma^+D^{*-}_{s0}$ decay rates are of order tens of MeV. In
particular, by taking $m_{\Theta_c}\simeq 3.1$~GeV as observed by
H1 collaboration \cite{H1}, we obtain $\Gamma(\Theta^0_c\to
pD^-)\simeq 3.1$~MeV, which is consistent with the observed width
of $\Gamma(\Theta_c)=12\pm3$~MeV \cite{H1}.
Taking $m_{\Xi^{--}_{3/2}}=1862\pm 2$~MeV as measured by NA49
\cite{NA49}, we obtain $\Gamma(\Xi^{--}_{3/2}\to\Sigma^-
K^-)\simeq 1.07$~MeV and
$\Gamma(\Xi^{--}_{3/2}\to\Xi^-\pi^-)\simeq 1.13$~MeV, which are
again consistent with the observed width of
$\Gamma(\Xi^{--}_{3/2})\leq 18$~MeV~\cite{NA49}. Our estimation
for the ratio
$\Gamma(\Xi^{--}_{3/2}\to\Xi^-\pi^-)/\Gamma(\Theta^+\to p
K^0)\simeq 2.2$ is several times smaller than that of \cite{Mehen}
but close to the estimate made in~\cite{Jaffe:2003ci}. Note that
the ratio $\Gamma(\Xi^{--}_{3/2}\to\Sigma^-
K^-)/\Gamma(\Xi^{--}_{3/2}\to\Xi^-\pi^-)\simeq 0.94$ is 50\%
larger than that obtained in \cite{Jaffe:2003ci} based solely on
the phase space consideration. The enhancement is due to the form
factor ratio $f^{\Xi_{3/2}\to
K}(m^2_\Sigma)/f^{\Xi_{3/2}\to\pi}(m^2_\Xi)=1.23$ (cf.
Table~\ref{tab:fg}) obtained in the LF calculation.

So far we have focused only on the strong decays of the
pentaquarks into an octet baryon and a pseudoscalar meson. For the
decay into a vector meson, it involves an additional unknown
tensor coupling which is calculable within our light-front
framework. Moreover, in the heavy quark limit $\P_Q\to M_Q\B$ and
$\P_Q\to M^*_Q\B$ are governed by the same strong coupling
constant. Indeed, heavy quark symmetry leads to the relation
$\Gamma(\P_Q\to M^*_Q\B)=3\Gamma(\P_Q\to M_Q\B)$ \cite{Manohar}.
Since $\Theta_c\to D^{*-}p$ has been observed by H1 \cite{H1}, it
will be interesting to measure the rate of $\Theta_c\to D^{-}p$ to
test heavy quark symmetry. Our result $\Gamma(\Theta^0_c\to
pD^-)\simeq 3.1$~MeV will imply $\Gamma(\Theta^0_c\to pD^{*-})\sim
9$~MeV. With $\Gamma(\Theta_c^0\to \overline D^{(*)0} n)\simeq
\Gamma(\Theta_c^0\to D^{(*)-} p)$ we expect
$\Gamma(\Theta_c)\simeq 20$~GeV. This is in accordance with the
observed width of $\Gamma(\Theta_c^0)=12\pm3$~MeV \cite{H1}.

It is interesting to note that although the $\P_c\to D_{s0}^*$
form factor is smaller than the $\P_c\to D_s$ one by a factor of
2, the decay rate for the $D_{s0}^*$ production is comparable to
that for $D_s$. This can be understood from the parity
consideration. Since $\P_c$, $\cal B$, $D_{s0}^*$ are parity even,
the final state ${\cal B} D_{s0}^*$ in $\P_c$ decay can have a
$s$-wave configuration, while the ${\cal B} D_s$ state must be in
a $p$-wave or higher odd-wave configuration, whose rate is
suppressed near the threshold. However, such a suppression is
absent in the final state composed of an even-parity meson and an
even-parity baryon and hence the decay $\P_c\to {\cal B} D_{s0}^*$
can have a sizable decay rate even its transition form factor is
suppressed. This is the reason why we have
$\Gamma(\Sigma_{5c}^0\to p
D_{s0}^{*-})\simeq\Gamma(\Sigma^0_{5c}\to pD_{s}^-)$ and
$\Gamma(\Xi_{5c}^0\to \Sigma^+
D_{s0}^{*-})\simeq\Gamma(\Xi_{5c}^0\to \Sigma^+ D_{s}^-)$ (see
Fig.~\ref{fig:rate}), provided that these strong decays are
kinematically allowed. As this is closely related to the
even-parity nature of these pentaquarks, the ratio of
$\Gamma(\P_c\to{\cal B} D^*_{s0})/\Gamma(\P_c\to{\cal B} D_s)$
provides for a useful way for verifying the parity of the charmed
pentaquark. For example, a completely opposite pattern --
$\Gamma(\Sigma_{5c}\to p D_{s0}^{*-})\ll\Gamma(\Sigma_{5c}\to
pD_{s})$ and $\Gamma(\Xi_{5c}\to \Sigma
D_{s0}^{*-})\ll\Gamma(\Xi_{5c}\to \Sigma D_{s})$ -- is expected
for odd-parity pentaquarks $\Sigma_{5c}$ and $\Xi_{5c}$.
It should be remarked that $m_{D_{s0}^*}\simeq
2.317$~GeV~\cite{BaBarDs,BelleDs} is substantially smaller than
expected from the quark model and hence the $D_{s0}^*{\cal B}$
threshold is close to the $D_s{\cal B}$ one, rendering the
production of the former easier than naive anticipation.

Finally, it is worth commenting that $\P_c$ can be produced in $B$
decays via $B\to\P_c\bar{\cal B}$ such as $B^+\to
\Theta_c^0\bar\Delta^+$ and $B^0\to\Theta_c^0\bar p\pi^+$
\cite{Rosner,Browder}. Theoretically, it is difficult to estimate
their branching ratios. Nevertheless, the measured branching
ratios by Belle for charmful baryonic $B$ decays \cite{Belle},
$\B(\ov B^0\to\Lambda_c^+\bar
p)=(2.2^{+0.6}_{-0.5}\pm0.3\pm0.6)\times 10^{-5}$ and
$\B(B^-\to\Lambda_c^+\bar
p\pi^-)=(1.87^{+0.43}_{-0.40}\pm0.28\pm0.49)\times 10^{-4}$,
provide some useful cue. Since a production of the pentaquark
needs one more pair of $q\bar q$ compared to the normal baryon, it
is plausible to expect that the branching ratios of $B^+\to
\Theta_c^0\bar\Delta^+$ and $B^0\to\Theta_c^0\bar p\pi^+$ are at
most of order $10^{-6}$ and $10^{-5}$, respectively. Hence, they
may be barely reachable at $B$ factories. Nervertheless, one can
search for $\P_c$ through $B\to \P_c\bar{\cal B'}\to (D {\cal
B})\bar{\cal B'},\,(D_s {\cal B})\bar{\cal B'},\,(D_{s0}^* {\cal
B})\bar{\cal B'}$ decays.

\section{conclusions}

Assuming the two diquark structure for the pentaquark as advocated
in the Jaffe-Wilczek model, we study the strong decays of
pentaquark baryons using the light-front approach in conjunction
with the spectator approximation. The main conclusions are as
follows.
 \begin{enumerate}
 \item
In the Jaffe-Wilczek model, the diquark pairs in the light
antidecuplet and heavy antisextet pentaquark baryons are in a
$p$-wave configuration. To describe their strong decays, the two
diquarks must interact to produce an ordinary baryon with a
$s$-wave quark configuration. This phenomenon has been modelled in
the present work by applying a local operator ${\cal O}_{\rm eff}$
for the $\phi\phi\to{\cal B}q$ transition. With a reasonable (and
unsuppressed) strong coupling of ${\cal O}_{\rm eff}$ we see that
the mismatch in the orbital angular momentum configuration is the
key physical reason for the narrowness of the pentaquark decay
width.
 \item
Treating the subprocess $\phi\phi\to{\cal B}q$ to be approximately
universal as suggested by the spectator picture, we estimate the
strong decays $\Xi^{--}_{3/2}\to\Xi^-\pi^-,\Sigma^-K^-$,
$\Theta^0_c\to pD^-$, $\Sigma_{5c}\to p D^-_s, pD^{*-}_{s0}$ and
$\Xi^0_{5c}\to \Sigma^+ D^-_s,pD^{*-}_{s0}$ by normalizing to the
$\Theta^+$ width. We find that
$\Xi^{--}_{3/2}\to\Xi^-\pi^-,\Sigma^-K^-$, $\Sigma_{5c}\to p
D^-_s, pD^{*-}_{s0}$ and $\Theta^0_c\to pD^-$decay rates are of
the order of a few MeV, while  and $\Xi^0_{5c}\to \Sigma^+ D^-_s,
\Sigma^+D^{*-}_{s0}$ decay rates are of order tens of MeV. If we
take $m_{\Xi^{--}_{3/2}}=1862\pm 2$~MeV as observed by NA49
\cite{NA49}, we have
$\Gamma(\Xi^{--}_{3/2}\to\Xi^-\pi^-)/\Gamma(\Theta^+\to
pK^0)\simeq 2.2$ which is consistent with the observed width of
$\Gamma(\Xi^{--}_{3/2})\leq 18$~MeV~\cite{NA49}.
 \item
Since the mass of the scalar meson $D_{s0}^{*-}$ is observed to be
lighter than expected, we also study $\P_c\to D_{s0}^*{\cal B}$
decays in addition to $\P_c\to D_{s}{\cal B}$ decays. The former
modes are enhanced (or unsuppressed) due to the even-parity nature
of $\P_c,{\cal B}$ and $D_{s0}^*$. In particular, the experimental
study of the ratio of $\Gamma(\P_c\to{\cal B}
D^*_{s0})/\Gamma(\P_c\to{\cal B} D_s)$ could be very useful for
verifying the parity of the sextet charmed pentaquark $\P_c$. It
is expected to be  of order unity for an even parity $\P_c$ and
much less than one for an odd parity one.
 \item
We also pointed out the possibility to search for $\P_c$ through
$B\to \P_c\bar{\cal B'}\to (D {\cal B})\bar{\cal B'},\,(D_s {\cal
B})\bar{\cal B'},\,(D_{s0}^* {\cal B})\bar{\cal B'}$ decays.
 \end{enumerate}

 \vskip 2.5cm \acknowledgments
 We are grateful to Hsiang-nan Li for valuable discussions.
This research was supported in part by the National Science
Council of R.O.C. under Grant Nos. NSC92-2112-M-001-016 and
NSC92-2811-M-001-054.

\appendix
\section{Clebsch-Gordan coefficients for Pentauqrks, Octet Baryons
and Mesons}

In this Appendix we specify the pentaquark, octet baryon and meson
$SU_f(3)$ quantum numbers. For anti-decuplet pentaquarks, we use
the totally symmetric tensor $P_{ijk}$ satisfying the
normalization condition $P_{ijk} P^{ijk}=1$
 \be
  P_{333}&=&\Theta^+,
  \non\\
  P_{133}=\frac{1}{\sqrt3} N^0_{\overline{10}},&&
  P_{233}=\frac{1}{\sqrt3} N^+_{\overline{10}},
  \non\\
  P_{113}=\frac{1}{\sqrt3} \Sigma^-_{\overline{10}},\quad
  P_{123}&=&\frac{1}{\sqrt6} \Sigma^0_{\overline{10}},\quad
  P_{223}=\frac{1}{\sqrt3} \Sigma^+_{\overline{10}},
  \non\\
  P_{111}=\Xi^{--}_{3/2},\quad
  P_{112}=\Xi^{-}_{3/2},&&
  P_{122}=\Xi^{0}_{3/2},\quad
  P_{222}=\Xi^{+}_{3/2}.
  \label{eq:Pijk}
 \en
Note that $(F_{L=1})_{ijk}=P_{ijk}$.
Anti-sextet heavy pentaquarks are described by the  totally
symmetric tensor $(\P_Q)_{ij}$. In the case of charm pentaquarks,
we have:
 \be
  (\P_c)_{33}&=&\Theta^0_c,
  \non\\
  (\P_c)_{13}=\frac{1}{\sqrt2} \Sigma^-_{5c},&&
  (\P_c)_{23}=\frac{1}{\sqrt2} \Sigma^0_{5c},
  \non\\
  (\P_c)_{11}=\Xi^{--}_{5c},\quad
  (\P_c)_{12}&=&\frac{1}{\sqrt2} \Xi^-_{5c},\quad
  (\P_c)_{22}= \Xi^0_{5c}.
  \label{eq:PQij}
  \en
The $SU_f(3)$ structure of octet baryons and mesons are
represented by ${\cal B}=T$ and $M$:
 \be
 {\cal B}= T=\left(
 \begin{array}{ccc}
 {{\Sigma^0}\over\sqrt2}+{{\Lambda}\over\sqrt6}
       &{\Sigma^+}
       &{p}
       \\
 {\Sigma^-}
       &-{{\Sigma^0}\over\sqrt2}+{{\Lambda}\over\sqrt6}
       &{n}
       \\
 {\Xi^-}
       &{\Xi^0}
       &-\sqrt{2\over3}{\Lambda}
 \end{array}
 \right),
 \quad
 M= \left(
 \begin{array}{ccc}
 {{\pi^0}\over\sqrt2}+{{\eta_8}\over\sqrt6}
       &{\pi^+}
       &{K^+}
       \\
 {\pi^-}
       &-{{\pi^0}\over\sqrt2}+{{\eta_8}\over\sqrt6}
       &{K^0}
       \\
 {K^-}
      &{\overline K {}^0}
      &-\sqrt{2\over3}{\eta_8}
 \end{array}
 \right).
 \label{eq:BM}
 \en
For example, for a final state $p$ and $K^0$ in the $\Theta^+$
decay, we need to use $T^3_1=p$, $M^3_2=K^0$.
Heavy mesons $(\overline D^0,D^-,D^-_s)$  transform like a triplet
$(u,d,s)$ under $SU_f(3)$.

\end{document}